\documentclass[traditabstract,onecolumn]{aamod} 
\usepackage{graphicx}
\usepackage{amsmath}
\usepackage{mathtools} 
\usepackage{booktabs} 
\usepackage{hyperref} 

\begin{document}

\title{A Bayesian method for the analysis of deterministic and stochastic time series}
\titlerunning{Bayesian time series analysis}
\author{C.A.L.\ Bailer-Jones}
\institute{Max Planck Institute for Astronomy, Heidelberg, Germany. Email: calj@mpia.de}
\date{Submitted 26 July 2012; revised 12 September 2012; accepted 13 September 2012}

\abstract{
I introduce a general, Bayesian method for modelling univariate time series data assumed to be drawn from a continuous, stochastic process.  The method accommodates arbitrary temporal sampling, and takes into account measurement uncertainties for arbitrary error models (not just Gaussian) on both the time and signal variables. Any model for the deterministic component of the variation of the signal with time is supported, as is any model of the stochastic component on the signal and time variables.  Models illustrated here are constant and sinusoidal models for the signal mean combined with a Gaussian stochastic component, as well as a purely stochastic model, the Ornstein--Uhlenbeck process.  The posterior probability distribution over model parameters is determined via Monte Carlo sampling. Models are compared using the ``cross-validation likelihood'', in which the posterior-averaged likelihood for different partitions of the data are combined.  In principle this is more robust to changes in the prior than is the evidence (the prior-averaged likelihood). The method is demonstrated by applying it to the light curves of 11 ultra cool dwarf stars, claimed by a previous study to show statistically significant variability.  This is reassessed here by calculating the cross-validation likelihood for various time series models, including a null hypothesis of no variability beyond the error bars. 10 of 11 light curves are confirmed as being significantly variable, and one of these seems to be periodic, with two plausible periods identified.  Another object is best described by the Ornstein--Uhlenbeck process, a conclusion which is obviously limited to the set of models actually tested.
}
\maketitle

\section{Introduction}

When confronted with a univariate time series, we are often interested in answering one or more of three questions. Which model best describes the data? What values of the parameters of this model best explain the data? What range of values does the model predict for the signal at some arbitrary time? These are questions of inference from data, and can be summarized as model comparison, parameter estimation and prediction, respectively. 

Probabilistic modelling provides a self-consistent and logical framework for answering these questions. In this article I introduce a general method for time series model comparison and parameter estimation.  The principle is straight forward.  The time series data comprise a set of measurements of the signal at various times, with measurement uncertainties generally in both signal and time.  We write down a parametrized model for the variation of the signal as a function of time. This could be a deterministic function or a stochastic model or, more generally, a combination of the two.  An example of a combined model is a sinusoidal variation of the mean of the signal on top of which is a Gaussian stochastic variation in the signal itself, which is {\em not} measurement noise. A purely stochastic model is one in which the expected signal evolves according to a random distribution, e.g.\ a random walk. 
Given this generative model and a noise model for the measurements, we then calculate the likelihood distribution of the data for different values of the model parameters. Rather than identifying just the single best fitting parameters, I use a Monte Carlo method to sample the posterior probability density function (PDF) over the model parameters. In addition to providing uncertainties on the inferred parameters, this also provides a measure of the goodness-of-fit of the overall model, in the form of the marginal likelihood (evidence), or the cross-validation likelihood (defined here). In this way we can identify the best overall model from a set, something which frequentist hypothesis testing can be notoriously bad at (e.g.\ Berger \& Sellke 1987, Kass \& Raftery 1996, Jaynes 2003, Christensen 2005, 
Bailer-Jones 2009).

There of course exist numerous time series analysis methods which attempt to answer one or more of the questions posed, so the reader may wonder why we need another one. For example, if we focus on periodic (Fourier) models, then we can calculate the power spectrum or periodogram in order to identify the most significant periods and to estimate the amplitudes of the components. If we work in the time domain, we could do least squares fitting of a parametrized model (e.g.\ Chatfield 1996, Brockwell \& Davis 2002).
However, many of these methods can only answer one of the posed questions, are limited to a restricted set of models or specific types of problems, do not take into account uncertainties in the signal and/or time, are limited to equally-spaced data,
do not provide uncertainty estimates on the parameters, or make other restrictive assumptions. The method introduced in the present work is quite general, and firmly embedded in a probabilistic approach to data modelling (see, e.g., von Toussaint 2011 for an introduction). This makes it powerful, but at the price of considerably higher computational cost. Yet in many applications this is a price we should be willing to pay for hard-won data, and should often be preferred to ad hoc, suboptimal recipes. 

The first two sections of this article -- occupying about a quarter of its length -- are dedicated to a complete description of the method: Section~\ref{sect:TSmethod} covers the method itself and the likelihood calculation, whereas the model comparison method described in section~\ref{sect:modcomp} is general. The mathematics here is relatively simple and intuitive.  More involved is the introduction of the Ornstein--Uhlenbeck process into the method. This, as well as the numerical approximations which allow the likelihood integrals to be evaluated more rapidly, are presented in the appendices. Section~\ref{sect:application} summarizes how to use the method. Most of the rest of the paper (about a third in total) is concerned with the application of the method, first to a simulated time series (section~\ref{sect:simulations}), and then to real astronomical data (section~\ref{sect:astroapp}). These are the light curves of 11 ultra cool dwarfs (low mass stars or brown dwarfs), which an earlier study has claimed show statistically significant variability. Although these data are used here primarily for demonstration purposes, this reanalysis of these data is astrophysically interesting, identifying a possible model and possible periods in two cases.  I summarize and conclude in section~\ref{sect:summary}.

The method developed here is related to the {\tt artmod} method introduced in 
Bailer-Jones (2011; hereafter CBJ11), which is a model for time-of-arrival time series. The present method extends this to model time series with noisy signal values at each measured time.

The notation used is summarized in Table~\ref{tab:notation}.

\begin{table}[!t]
\begin{center}
\caption{Primary notation \label{tab:notation}}\vspace*{1em}
\begin{tabular}{ll}
\toprule
symbol & definition \\
\midrule
$s_j$  &  measured time of $j^{\rm th}$ event \\
$\sigma_{s_j}$  &  standard deviation in $s_j$ \\
$t_j$   & (unknown) true time of the $j^{\rm th}$ event \\
$y_j$  &  measured signal of $j^{\rm th}$ event \\
$\sigma_{y_j}$  &  standard deviation in $y_j$ \\
$z_j$   & (unknown) true signal of the $j^{\rm th}$ event \\
$D_j$  &  $=(s_j, y_j)$ measurements for the $j^{\rm th}$ event\\
$\sigma_j$  &  $=(\sigma_{s_j}, \sigma_{y_j})$ estimated uncertainties in $D_j$ \\
$D$ & $=\{D_j\}$ set of measurements for $J$ events \\
$\sigma$ & $=\{\sigma_j\}$  estimated uncertainties in $D$ \\
$D_k$ & set of measurements for events in partition $k$ \\
$D_{-k}$ & set of measurements for all events not in in partition $k$ \\
$M$  &  time series model\\
$\theta$  &  $=(\theta_1, \theta_2, \theta_3)$, parameters of the time series model \\
$\eta(t ; \theta_1)$ & deterministic model of the expected true signal (TSMod1) \\
$\log$ & logarithm base 10 \\
${\cal N}(x ; \mu, V)$ & Gaussian in $x$ with mean $\mu$ and variance $V$\\
\bottomrule
\end{tabular}
\end{center}
\end{table}

\section{The time series method}\label{sect:TSmethod}

\subsection{Data and model definition}

\begin{figure}
\begin{center}
\includegraphics[width=0.5\textwidth]{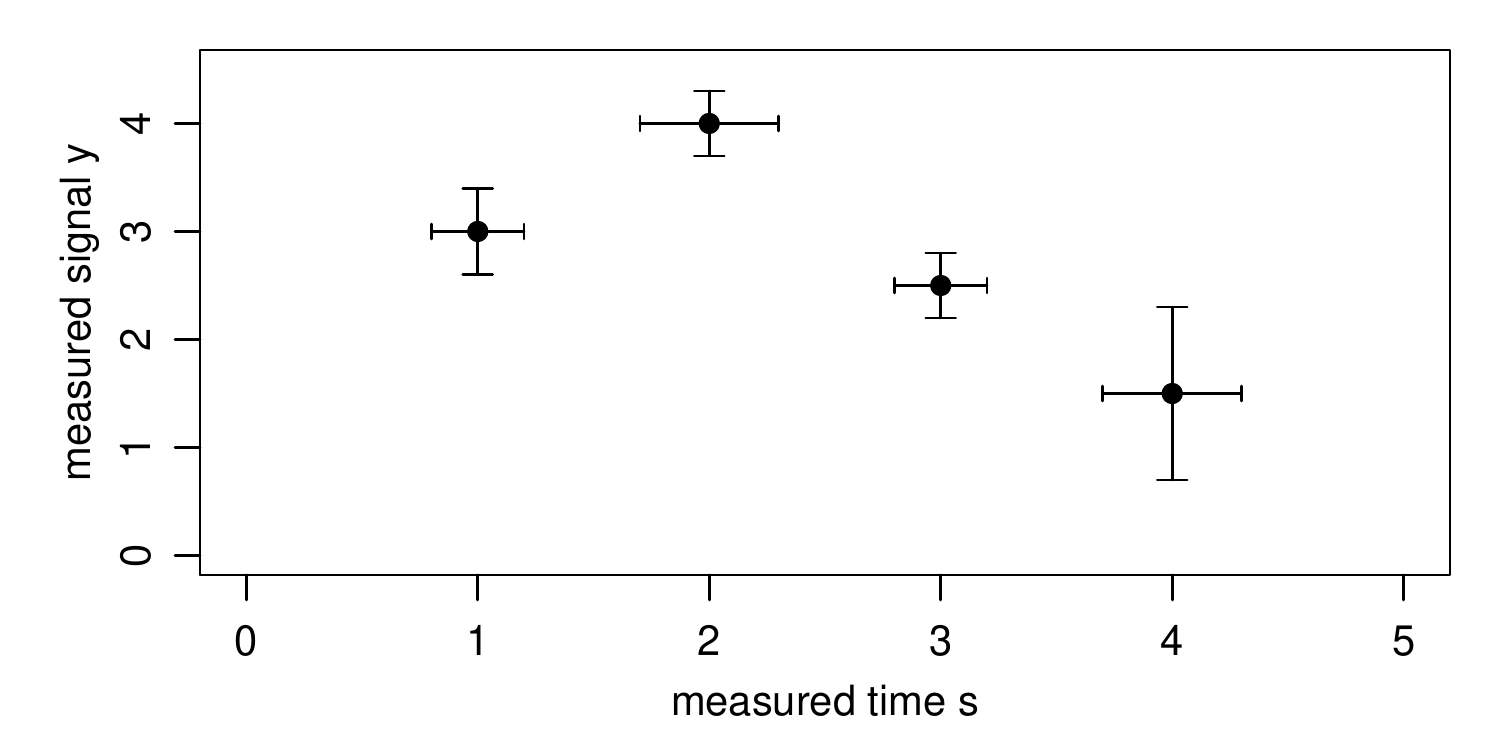}
\caption[]{Example of a measured data set with four events
\label{fig:dataplot}}
\end{center}
\end{figure}

We have a set of $J$ events, each defined by its time, $t$, and signal, $z$.
For each event $j$, our measurement of the time of the event, $t_j$, is $s_j$ with a standard deviation (estimated measurement uncertainty) $\sigma_{s_j}$, and our measurement of the signal of the event, $z_j$, is $y_j$ with a standard deviation (estimated measurement uncertainty) $\sigma_{y_j}$ (see Figure~\ref{fig:dataplot}). 
That is, $t_j$ and $z_j$ are the true, unknown values, not the measurements. 
Define $D_j=(s_j, y_j)$ and $\sigma_j=(\sigma_{s_j}, \sigma_{y_j})$.  The {\em measurement model} (or noise model) describes the probability of observing the measured values for a single event given the true values and the estimated uncertainties: it gives $P(D_j | t_j, z_j, \sigma_j)$. The $\sigma_j$ are considered fixed parameters of the measurement model, and the conditioning on the measurement model is implicit (because I do not want to compare measurement models in this work).

$M$ is a stochastic {\em time series model} with parameters $\theta$. It specifies $P(t_j, z_j | \theta, M)$, the probability of observing an event at time $t_j$ with signal $z_j$.

The goal is (1) to compare the posterior probability of different models $M$, and (2) to determine the posterior PDF over the model parameters for a given $M$. After describing the measurement and time series models in the next two subsections, I will then show how to combine them in order to calculate the likelihood.

\subsection{Measurement model}\label{sect:measmod}

If $t$ and $z$ have no bounds, or can be approximated as such, and the known measurement uncertainties are standard deviations, then an appropriate choice for the measurement model is a two-dimensional (2D) Gaussian in the variables $(s_j,y_j)$ for event $j$.
If we assume no covariance between the variables then this reduces to the product of two 1D Gaussians
\begin{equation}
P(D_j | t_j, z_j, \sigma_j) = \frac{1}{\sqrt{2\pi}\sigma_{s_j}} \, e^{-(s_j - t_j)^2/2\sigma_{s_j}^2}  \,\, \frac{1}{\sqrt{2\pi}\sigma_{y_j}} \, e^{-(y_j - z_j)^2/2\sigma_{y_j}^2} \ .
\label{eqn:measmod}
\end{equation}
(The two terms are normalized with respect to $s_j$ and $y_j$ respectively.)
If we had other information about the measurement, e.g.\ asymmetric error bars, strictly positive signals, or uncertainties which are not standard deviations, then we should adopt a more appropriate distribution.

\subsection{Time series model}\label{sect:tsmodel}

\begin{figure}
\begin{center}
\includegraphics[width=0.5\textwidth]{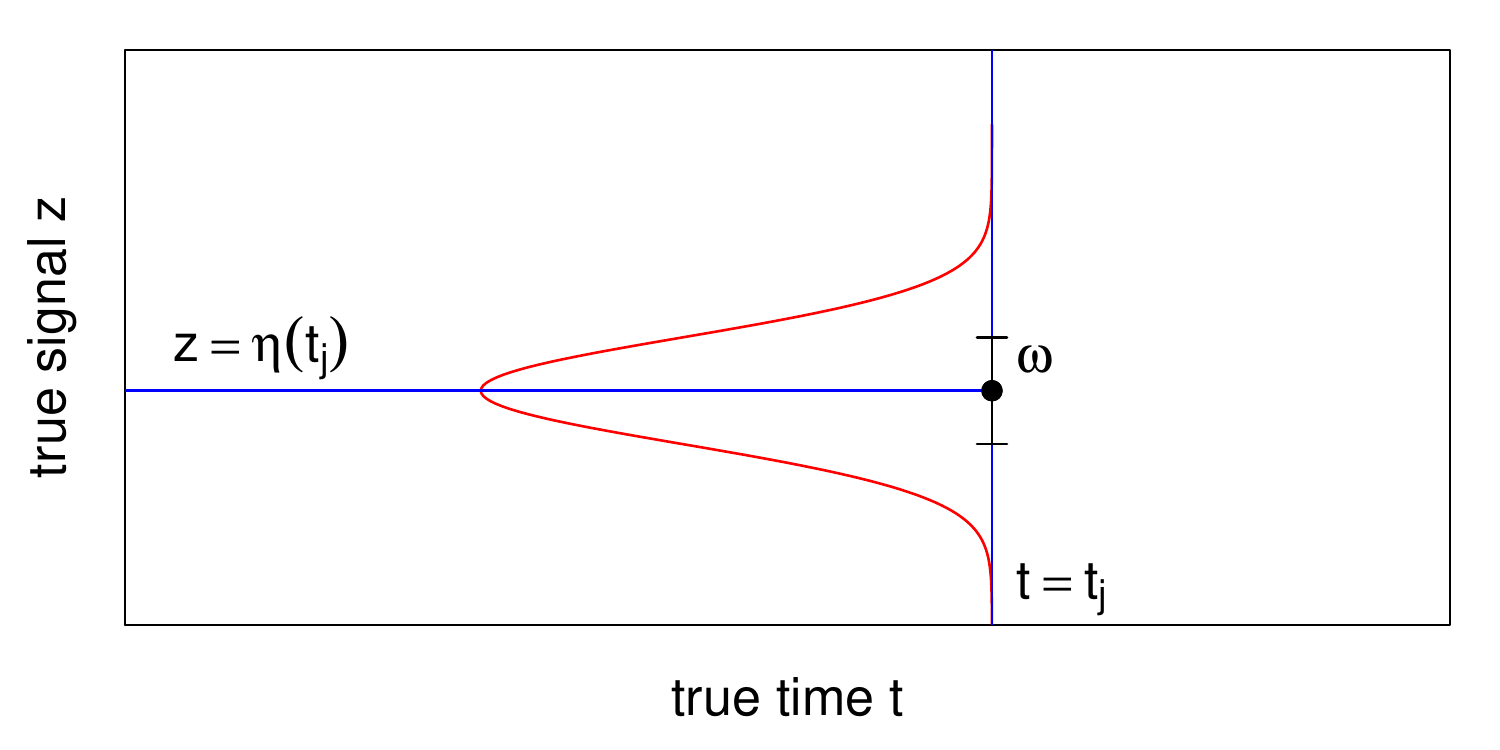}
\caption[]{Conceptual representation of the stochastic nature of the signal component of the time series model, $P(z_j | t_j, \theta^{\prime}, M)$ (in red) of the true signal, $z$, at a true time $t_j$ (here shown as a Gaussian).
\label{fig:pztplot}}
\end{center}
\end{figure}

Without loss of generality, the time series model can be written as the product of two stochastic components
\begin{equation}
P(t_j, z_j | \theta, M) \,=\, P(z_j | t_j, \theta, M) P(t_j | \theta, M) 
\label{eqn:tsmod}
\end{equation}
which I will refer to as the signal and time components respectively. For many processes it is appropriate to express the signal component using two independent subcomponents: the stochastic model itself and a deterministic function which defines the time-dependence of its mean.
This stochastic subcomponent describes the intrinsic variability of the true signal of the physical process at a given time, with the PDF $P(z_j | t_j, \theta^{\prime}, M)$. I refer to this as TSMod2. An example is a Gaussian
\begin{equation}
P(z_j | t_j, \theta^{\prime}, M) \,=\, \frac{1}{\sqrt{2\pi}\omega} \, e^{-(z_j - \eta[t_j])^2/2\omega^2}  
\label{eqn:tsmod2gaussian}
\end{equation}
where $\theta^{\prime} = (\eta, \omega$) are the parameters of the distribution: $\eta[t_j]$ is the expected true signal at true time $t_j$; $\omega$ is a parameter which reflects the degree of stochasticity in the process. This is illustrated schematically for a single point in Figure~\ref{fig:pztplot}. The Gaussian is just an example, and would be inappropriate if $z$ were a strictly non-negative quantity.

The relationship between the expected true signal and the true time is given by a deterministic function, $\eta(t ; \theta_1)$, where $\theta_1$ denotes another set of parameters. I refer to this deterministic subcomponent as TSMod1. A simple example is a single frequency sinusoid
\begin{equation}
\eta \,=\, \frac{a}{2} \cos[2\pi (\nu t + \phi)] + b
\label{eqn:tsmod1sinusoid}
\end{equation}
which has parameters $\theta_1 = (a, \nu, \phi, b)$, the amplitude, frequency, phase, and offset. Having parametrized the signal component of the time series model in this way, it is convenient to write $\theta^{\prime} = (\theta_1, \theta_2)$ in general, where $\theta_2 = \omega$ in the example of equation~\ref{eqn:tsmod2gaussian}.

The second component of the time series model in equation~\ref{eqn:tsmod} describes any intrinsic randomness in the time of the events which make up the physical process. This is represented by $P(t_j | \theta_3, M)$, which I refer to as TSMod3. 
If there is no variation in the probability when an event could occur, we should make this constant by using
a flat distribution in $t$ between the earliest possible and latest possible times, $T_1$ and $T_2$, 
\begin{equation}
P(t_j | \theta_3, M)  \,=\,  
\begin{dcases}
 \frac{1}{T_2 - T_1}  & \:{\rm if}~~ T_1  < t_j < T_2 \\
 0                            & \:{\rm otherwise}
\end{dcases}
\label{eqn:tsmod3uniform}
\end{equation}
where $\theta_3 = (T_1, T_2)$ are its parameters. This is generally appropriate to modelling light curves (for example), where there is no concept of intrinsically discrete events: the discreteness arises only because we take measurements at certain times.  There is then no sense in which the probability of an ``event occuring'' could vary. In contrast, when modelling a discrete process, such as the times and energies of large asteroid impacts (see CBJ11), we generally will have a time varying probability of an event occuring.

This three-subcomponent approach (TSMod1,2,3) to the time series model is conceptually a little complex, so let us consider what it means. We have a physical process in which the expected value of the signal varies with time in a deterministic manner. This is given by $\eta(t ; \theta_1)$, e.g.\ equation~\ref{eqn:tsmod1sinusoid}. At any given true time, the true signal of the process can vary due to intrinsic randomness in the process. This is described by
$P(z_j | t_j, \theta_1, \theta_2, M)$, an example of which is equation~\ref{eqn:tsmod2gaussian}. Finally, while the mean of the process signal is considered to vary continuously in time, there may be a time varying probability that an event could occur at all (e.g.\ an asteroid impact). This is described by $P(t_j | \theta_3, M)$. 
The stochasticity in the time series model has nothing to do with measurement noise. It is intrinsic to the process.

This description of the signal component as a stochastic model with a time-independent variance and a (deterministic) time-dependence for the mean we might refer to as a {\em partially} stochastic process.  A {\em fully} stochastic process, in contrast, is one in which {\em all} the parameters of the PDF $P(z_j | t_j, \theta, M)$ have a time-dependence, in which case this decomposition of the signal component into TSMod1 and TSMod2 is not possible.
An example is the Ornstein--Uhlenbeck process, which will be used in this work. It is described in appendix~\ref{sect:fullystochastic}.

The overall time series model is the combination of these three subcomponents
\begin{equation}
P(t_j, z_j | \theta, M) = P(z_j | t_j, \theta_1, \theta_2, M) P(t_j | \theta_3, M)
\end{equation}
where $\theta = (\theta_1, \theta_2, \theta_3)$.  For the cases shown above, this model has seven parameters, $\theta = (a, \nu, \phi, b, \omega, T_1, T_2)$, although probably we would fix $(T_1, T_2)$ based on inspection of the time range of the data.

Later in section~\ref{sect:prior} we will look at the specific models and their parametrizations as used in this paper.

\subsection{Likelihood}

The probability of observing data $D_j$ from time series model $M$ with parameters $\theta$ when the uncertainties are $\sigma_j$, is
$P(D_j | \sigma_j, \theta, M)$, the {\em event likelihood}. This is obtained by marginalizing over the true, unknown event time and signal
\begin{alignat}{2}
P(D_j | \sigma_j, \theta, M) \,&=\, \int_{t_j, z_j} P(D_j, t_j, z_j | \sigma_j, \theta, M) \; dt_j dz_j \nonumber \\ 
                            \,&=\, \int_{t_j, z_j} P(D_j | t_j, z_j, \sigma_j, \theta, M) P(t_j, z_j | \sigma_j, \theta, M) \; dt_j dz_j \nonumber \\ 
                            \,&=\, \int_{t_j, z_j} \underbrace{P(D_j | t_j, z_j, \sigma_j)}_\text{Measurement model} \: \underbrace{P(t_j, z_j | \theta, M)}_\text{Time series model} \; dt_j dz_j
\label{eqn:eventlike}
\end{alignat}
where the time series model and its parameters drop out of the first term because $D_j$ is independent of this once conditioned on the true variables, and the measurement model (via $\sigma_j$) drops out of the the second term because it has nothing to do with the predictions of the time series model. For specific, but common, situations, this 2D integral can be approximated by a 1D integral or even a function evaluation (see appendix~\ref{sect:simplifications}).

If we have a set of $J$ events for which the ages and signals have been estimated independently of one another, then the probability of 
observing these data $D=\{D_j\}$, the {\em likelihood}, is
\begin{equation}
P(D | \sigma, \theta, M) \,=\, \prod_j P(D_j | \sigma_j, \theta, M)
\label{eqn:likelihood}
\end{equation}
where $\sigma=\{\sigma_j\}$.

\section{Model comparison}\label{sect:modcomp}

\subsection{Evidence}\label{sect:evidence}

In order to compare different models, $M$, we would like to know $P(M|D,\sigma)$, the model posterior probability.  We can use Bayes' theorem to write this down in terms of the evidence, $P(D | \sigma, M)$.  This is the probability of getting the observed data from model $M$, regardless of the specific values of the model parameters.  Adopting equal prior probabilities, $P(M)$, for different models, the evidence is the appropriate quantity to use to compare models.  One may be tempted to use instead the maximum value of the likelihood for model comparison, but this is wrong, because it will just favour increasingly complex models, as these can increasingly overfit the data. For more discussion of this point see, for example, MacKay (2003) or CBJ11.

The {\em evidence}, $E$, is obtained by marginalizing the likelihood over the parameter prior probability distribution, $P(\theta | M)$.
\begin{alignat}{2}
E \,\equiv\, P(D | \sigma, M)  \,&=\,  \int_{\theta} P(D, \theta | \sigma, M) \; d\theta \nonumber \\ 
                                              \,&=\,  \int_{\theta} \underbrace{P(D | \sigma, \theta, M)}_\text{likelihood} \; \underbrace{P(\theta | M)}_\text{prior} \; d\theta \ .
\label{eqn:evidence}
\end{alignat}
Note that the evidence is conditioned on both the measurement model (via $\sigma$) and the time series model, $M$.  For a given set of data, we calculate this evidence for the different models we wish to compare, each parametrized by $\theta$.  The parameter prior, $P(\theta | M)$, encapsulates our knowledge of the prior plausibility of different parameters. (This is independent of $\sigma$, which is why it was removed in the above equation.)  As the evidence has an uninterpretable scale, we usually examine the ratio of the evidences of two models, the Bayes factor.

We evaluate the above integral using a Monte Carlo approximation
\begin{equation}
E \:\approx\: \frac{1}{N}\sum_{n=1}^{n=N}  P(D | \sigma, \theta_n, M)
\label{eqn:montecarlo}
\end{equation}
where the parameter samples, $\{\theta_n\}$, have been drawn from the prior, $P(\theta | M)$. Often this prior is a product of simple, 1D functions (e.g.\ Gaussian or Gamma PDFs), so it is easy to sample from without having to employ more sophisticated methods.

\subsection{Cross validation likelihood}\label{sect:CV}

The evidence is often sensitive to the parameter prior PDF. For example, in a single-parameter model, if the likelihood were constant over the range $0<\theta<1$ but zero outside this, then the evidence calculated using a prior uniform over $0<\theta<2$ would be half that calculated using a prior uniform over $0<\theta<1$. In a model with $p$ such parameters the factor would be $2^{-p}$. If we had no reason to limit the prior range, then the evidence would be of limited use in this example. 
Conversely, in cases where the parameters have a physical interpretation and/or where we have reasonable prior information, then we may be able to justify a reasonable choice for the prior. But in any case we should always explore the sensitivity of the evidence to ``fair'' changes in the prior. A fair change is one which we have no reason not to make. For example, if there were no reason to prefer a prior which is uniform over frequency rather than period, then this would be a fair change. (See CBJ11 for an illustration of this on real data.) If the evidence changes enough to alter significantly the Bayes factors when making fair changes, then the evidence is over-sensitive to the choice of prior, making it impossible to draw robust conclusions without additional information.

In such situations we might resort to one of the many ``information criteria'' which have been defined, such as the Akaike Information Criterion (AIC) or Bayesian Information Criterion (BIC) (e.g.\ Kadane \& Lazar 2004) or the Deviance Information Criterion (DIC) (Spiegelhalter et al.\ 2002).  The advantage of these is that they are simpler and quicker to calculate. 
But they all make (possibly unreasonable) assumptions regarding how to represent the complexity of the model, and all have been criticized in the literature.

An alternative approach is a form of K-fold cross validation (CV). We split the data set ($J$ events) into $K$ disjoint partitions, where $K \leq J$. Denote the data in the $k^{\rm th}$ partition as $D_k$ and its complement as $D_{-k}$. The idea is to calculate the likelihood of $D_k$ using $D_{-k}$, without having an additional dependence on a specific choice of model parameters. That is, we want $P(D_k | D_{-k}, \sigma, M)$, which tells us how well, in model $M$, some of the data are predicted using the other data.
Combining these likelihoods for all $K$ partitions gives an overall measure of the fit of the model. By marginalization
\begin{alignat}{2}
P(D_k | D_{-k}, \sigma, M) \,&=\, \int_\theta P(D_k | D_{-k}, \sigma, \theta, M) P(\theta | D_{-k}, \sigma, M) d\theta \nonumber \\ 
                            \,&=\, \int_\theta \underbrace{P(D_k | \sigma_{k}, \theta, M)}_\text{likelihood} \underbrace{P(\theta | D_{-k}, \sigma_{-k}, M)}_\text{posterior} d\theta
\label{eqn:partlike}
\end{alignat}
where $D_{-k}$ drops out of the first term because the model predictions are independent of these data when $\theta$ is
specified. ($\sigma_{-k}$ and $\sigma_{k}$ drop out of the first and second terms, respectively, also for reasons of independence.) (Cf.\ equation 10 of Vehtari \& Lampinen 2002.) If we draw a sample $\{\theta_n\}$ of size $N$ from the posterior $P(\theta | D_{-k}, \sigma_{-k}, M)$, then the Monte Carlo approximation of this integral is
\begin{equation}
L_k \,\equiv\, P(D_k | D_{-k}, \sigma, M) \,\approx\, \frac{1}{N} \sum_{n=1}^{n=N} P(D_k | \sigma_{k}, \theta_n, M)
\label{eqn:partlike2}
\end{equation}
the mean of the likelihood of the data in partition $k$. I will call $L_k$ the {\em partition likelihood}. Note that here the posterior is sampled using the data $D_{-k}$ only.

Because $L_k$ is the product of event likelihoods, it scales multiplicatively with the number of events in partition $k$. An appropriate combination of the partition likelihoods over all partitions is therefore their product
\begin{equation}
L_{\rm CV} \,=\, \prod_{k=1}^{k=K} L_k  \;\;\;\; {\rm or} \;\;\;\; \log L_{\rm CV} \,=\, \sum_k \log L_k
\label{eqn:kfoldcvlike}
\end{equation}
which I call the {\em K-fold cross validation likelihood}, for $1 \leq K \leq J$.
If  $K>1$ and $K<J$ then its value will depend on the choice of partitions. 
If $K=J$ there is one event per partition (a unique choice). This is {\em leave-one-out CV (LOO-CV)}, the likelihood for which I will denote with $L_{\rm LOO-CV}$.
If $K=1$, we just use all of the data to calculate both the likelihood and the posterior.
This is not a very correct measure of goodness-of-fit, however, because it uses all of the data both to draw the posterior samples and to calculate the likelihood. 

The posterior PDF required in equation~\ref{eqn:partlike} is given by Bayes' theorem. It is sufficient to use the unnormalized posterior (as indeed we must, because the normalization term is the evidence), which is
\begin{equation}
P(\theta | D_{-k}, \sigma_{-k}, M) \,\propto\, P(D_{-k} | \sigma_{-k}, \theta, M) P(\theta | M)
\label{eqn:postPDF}
\end{equation}
i.e.\ the product of the likelihood and the prior. 
$L_{\rm CV}$ therefore still depends on the choice of prior (discussed in section~\ref{sect:prior}). However, the likelihood will often dominate the prior (unless the data are very indeterminate), in which case $L_{\rm CV}$ will be less sensitive to the prior than is the evidence.

There is a close relationship between the partition likelihood and the evidence.
Whereas the evidence involves integrating the likelihood (for $D$) over the {\em prior} (equation~\ref{eqn:evidence}), the partition likelihood involves integrating the likelihood (for $D_k$) over the {\em posterior} (for $D_{-k}$) (equation~\ref{eqn:partlike}). This is like using $D_{-k}$ to build a new prior from ``previous'' data.
We can use the product rule to write the partition likelihood as
\begin{equation}
L_k \,\equiv\, P(D_k | D_{-k}, \sigma_k, M) \,=\, \frac{P(D | \sigma, M)}{P(D_{-k} | \sigma_{-k}, M)} 
\end{equation}
which shows that it is equal to the ratio of the evidence calculated over all the data to the evidence calculated on the subset of the data used in the posterior sampling. As the same prior PDF enters into both terms, it will, in some vague sense, ``cancel'' out, although I stress that there is still a prior dependence.

It is important to realize that the model complexity is taken into account by the model comparison with the K-fold CV likelihood (and therefore the LOO-CV likelihood), just as it is with the Bayesian evidence.  That is, more complex models are not penalized simply on account of having more parameters. It is, as usual, the prior plausibility of the model which counts.

\subsection{Parameter priors}\label{sect:prior}

The model measures mentioned -- the evidence, the K-fold CV likelihood, also the DIC -- are calculated by averaging the likelihood over the model parameter space.  This parameter space must therefore be sampled, and this requires that we specify a prior PDF, $P(\theta|M)$, over these.

We invariably have some information about values of the parameters, such as bounds or plausible values. For example, standard deviations, frequencies and amplitudes cannot be negative, and a phase (as defined here) must lie be between 0 and 1. Non-negative quantities are common, and for these I adopt the gamma distribution in the applications which follow. This is characterized by two parameters, shape and scale (both positive). The mean of the gamma distribution is shape times scale, and its variance is the mean times scale.

The different components of the time series models used in the applications, along with the prior distributions over their parameters, are shown in Table~\ref{tab:tsmodels}.

\begin{table}
\begin{center}
\caption{Time series model components with parameters $\theta$ used in this work. The penultimate column shows the prior PDFs over these parameters, which themselves have (hyper)parameters $\alpha$. These prior PDFs respect the limits on $\theta$ listed in the final column. 
$U()$ is the uniform distribution between 0 and 1, so has no free parameters.
Note that $a$ in the Sin model is the peak-to-peak amplitude of the sinusoid. The uniform model for TSMod3 is used throughout (equation~\ref{eqn:tsmod3uniform}).
\label{tab:tsmodels}}\vspace*{1em}
\begin{tabular}{lp{4.5cm}lll}
\toprule
name               & function & $\theta$   & prior PDF, $P(\theta ; \alpha)$ & $\theta_{\rm min}, \theta_{\rm max}$\\  
\midrule
\multicolumn{5}{l}{{\bf TSMod1}} \\
Off    & $b$        & $b$    & ${\cal N}(b ; {\rm mean}, {\rm sd})$                & $-\infty,\infty$ \vspace*{0.25em}  \\
Sin    &  ${\displaystyle \frac{a}{2}} \cos[2\pi (\nu t + \phi)] $ & $a$     & ${\rm Gamma}(a ; {\rm shape}, {\rm scale})$    & $0,\infty$  \vspace*{0.25em}  \\
                        &                                                                                          & $\nu$  & ${\rm Gamma}(\nu ; {\rm shape}, {\rm scale})$ & $0,\infty$  \vspace*{0.25em}  \\
                        &                                                                                          & $\phi$ & $U(\phi)$ & $0,1$ \vspace*{0.25em}  \\
\midrule
\multicolumn{5}{l}{{\bf TSMod2}} \\
Stoch   &  $\frac{1}{\sqrt{2\pi}\omega} \, e^{- (z - \eta[t ; \theta_1])^2/2 \omega^2}$    & $\omega$     & ${\rm Gamma}(\omega ; {\rm shape}, {\rm scale})$ & $0,\infty$  \vspace*{0.25em}  \\
OUprocess   &  see appendix~\ref{sect:fullystochastic} &  $\tau$    & ${\rm Gamma}(\tau ; {\rm shape}, {\rm scale})$ & $0,\infty$  \vspace*{0.25em}  \\
                        &                                                                                          & $c$  & ${\rm Gamma}(c ; {\rm shape}, {\rm scale})$ & $0,\infty$  \vspace*{0.25em}  \\
                        &                                                                                          & $b$  & ${\cal N}(b ; {\rm mean}, {\rm sd})$ & $-\infty,\infty$  \vspace*{0.25em}  \\
                        &                                                                                          & $\mu[z_1]$  & ${\cal N}(\mu[z_1] ; {\rm mean}, {\rm sd})$ & $-\infty,\infty$  \vspace*{0.25em}  \\
                        &                                                                                          & $\sqrt{V[z_1]}$  & ${\rm Gamma}(\sqrt{V[z_1]} ; {\rm shape}, {\rm scale})$ & $0,\infty$ \\
\bottomrule
\end{tabular}
\newline
\end{center}
\end{table}

\begin{figure}
\begin{center}
\includegraphics[width=0.5\textwidth]{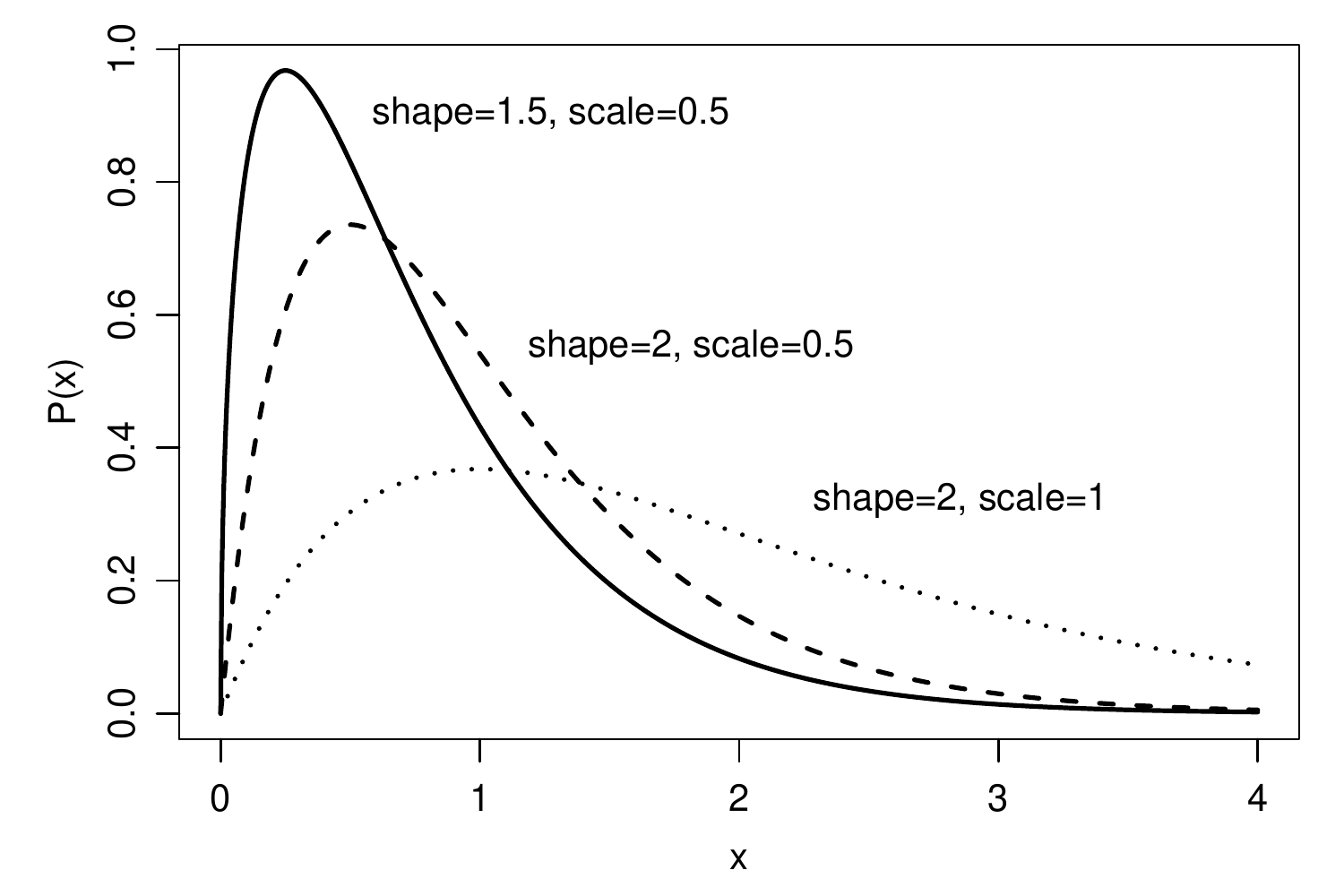}
\caption[]{Three examples of the gamma PDF, used as the prior for non-negative model parameters. The solid line, with
shape=1.5, scale=0.5, is used as the prior PDF over frequency in units of inverse hours.
\label{fig:gammaPDF}}
\end{center}
\end{figure}

We have to assign values for the (hyper)parameters, $\alpha$, of the prior PDFs. Although we rarely have sufficient knowledge to specify these precisely, we can use our knowledge of the problem and the general scale of the data to assign them. I adopt the following procedure for assigning what I call the {\em canonical priors}, appropriate for the data which will be analysed in section~\ref{sect:astroapp}.
Some parameters are set according to the standard deviation of the signal values, $\varsigma_y = \sqrt{\frac{1}{J-1}\sum_j (y_j-\overline{y_j})^2}$, where $\overline{y_j}$ is the mean signal.
\begin{itemize}
\item For the Off model (parameter $b$), I use a Gaussian with zero mean and standard deviation 1--2 times $\varsigma_y$. The exact value is determined by visual inspection of the light curve.
\item  For the Sin model, I use a gamma prior on the frequency, $\nu$, with shape=1.5 and scale=0.5 (Fig.~\ref{fig:gammaPDF}). This assigns significant prior probability to a broad range of frequencies believed to be plausible based on knowledge of the problem, the temporal sampling, and the total span of the light curves. 
For the amplitude, $a$, I use a gamma prior with shape=2 and scale 1--3 times $\varsigma_y$. The prior over the phase is uniform.
\item For the Stoch model (parameter $\omega$), I use a gamma prior with shape=2 and scale 1--2 times $\varsigma_y$.
\item For the OU process, I use a gamma prior on both $\tau$ and $c$ with shape=1.5. $\tau$ is a decay timescale, so I set its scale parameter to one quarter of the duration of the time series. The long-term variance of the OU process is $c\tau/2$. Equating this to $\varsigma_y^2$, I therefore set the scale of the diffusion coefficient, $c$, to be $2\varsigma_y^2/\tau$.
The parameters $b$ and $\mu[z_1]$ are both assigned Gaussian priors with a standard deviation equal to $\varsigma_y$.
The mean of the former is set to zero, and the mean of the latter to $y_1$, the signal value of the first data point.
The final parameter, $\sqrt{V[z_1]}$, a standard deviation, is assigned a gamma distribution with shape=1.5 and scale=$\varsigma_y$.
\end{itemize}
This scheme of ``data-based'' priors was arrived at after some experimentation, and generally the calculating LOO-CV likelihoods are robust to small changes in the priors (as demonstrated later).

\subsection{Markov Chain Monte Carlo (MCMC)}\label{sect:MCMC}

For sampling the posterior I use the standard Metropolis algorithm with a Gaussian sampling distribution with diagonal covariance matrix. Those model parameters which do not naturally have an infinite range are transformed in order to be commensurate with Gaussian sampling: parameters with a range zero to infinity (such as frequency) are logarithmically transformed; phase (which has a range 0-1) is transformed using the logit (inverse sigmoid) function. 
The standard deviations of the sampling covariance matrix are set to fixed, relatively small values, typically 0.05--0.1 for the logarithmically transformed parameters (these are then scale factors).
A consequence of this scaling is that the parameter can never exactly reach the extreme values (zero for the log transformation), but this is not necessarily a disadvantage. 
I experimented with instead using a circular transformation rather than logit for the phase
parameter (by taking $\phi\ {\rm mod}\ 1$). While this has an affect on the posterior phase distribution, it barely changed the resulting model average likelihoods.


\section{Practical application}\label{sect:application}

Given a data set and a time series model we wish to evaluate, the procedure for applying the model is as follows:
(1) define the (hyper)parameters of the prior parameter PDFs, as well as the standard deviations of the MCMC sampling PDF and its initial values; (2) select a partitioning of the data (normally we will use LOO-CV, so the choice is unique); (3) for each partition of the data, use MCMC to sample the posterior PDF, retaining the value of the likelihood at each parameter sample. Average these likelihoods to get the partition likelihood (equation~\ref{eqn:partlike2}); (4) sum the logarithms of the partition likelihoods to get the K-fold CV log likelihood (equation~\ref{eqn:kfoldcvlike}).
Note that each partition provides a posterior PDF, which we could plot and summarize.
In order to calculate the evidence for a model (equation~\ref{eqn:evidence}), then after step (1) we sample the prior PDF and use equation~\ref{eqn:montecarlo}.

In the applications in this article I adopt the uniform model for the temporal component of the time series model, TSMod3 (equation~\ref{eqn:tsmod3uniform}). The two parameters of this are fixed to the start and end points of the measured light curve to include all of the data. (The exact values are otherwise irrelevant, as they are the same for all models for a given light curve. 
This effectively removes TSMod3 from the model.)
I also assume that the signal component of the measurement model is a Gaussian. I further assume that the uncertainties on the times are small compared to the time scale on which the time series model varies. This allows us to replace the 2D integration in the expression for the event likelihood (equation~\ref{eqn:eventlike}) with an analytic expression, as shown in section~\ref{sect:errsmall}. This results in significantly reduced computational times.

In section~\ref{sect:errsmall} I define the {\em no-model}, the model which assumes that the data
are just Gaussian variations -- with standard deviation given by the error bars -- about the mean of the data.  As this model has no parameters, its likelihood, $L^{\rm NM}$, is equal to its LOO-CV likelihood and its evidence. This is therefore a convenient
 baseline against which to compare all other models, so in the text and tables I report the LOO-CV likelihood/evidence for all models relative to this, i.e.\ $\log L_{\rm LOO-CV} - \log L^{\rm NM}$ and $E - \log L^{\rm NM}$ (the latter is the logarithm of the Bayes factor).

Once we have calculated these quantities for a number of models, we need to compare them.  It is somewhat arbitrary how large the difference in the log likelihoods must be before we bother discussing them.  Clearly very small differences are not ``significant'', as small changes in the priors would produce ``acceptable'' changes in the likelihoods. Here I identify two models as being ``significantly different'' if their log (base 10) likelihoods differ by more than 1. I use this term merely for the sake of identifying which differences are worth discussing.

\section{Application to simulated time series}\label{sect:simulations}


\begin{figure}
\begin{center}
\includegraphics[width=0.5\textwidth]{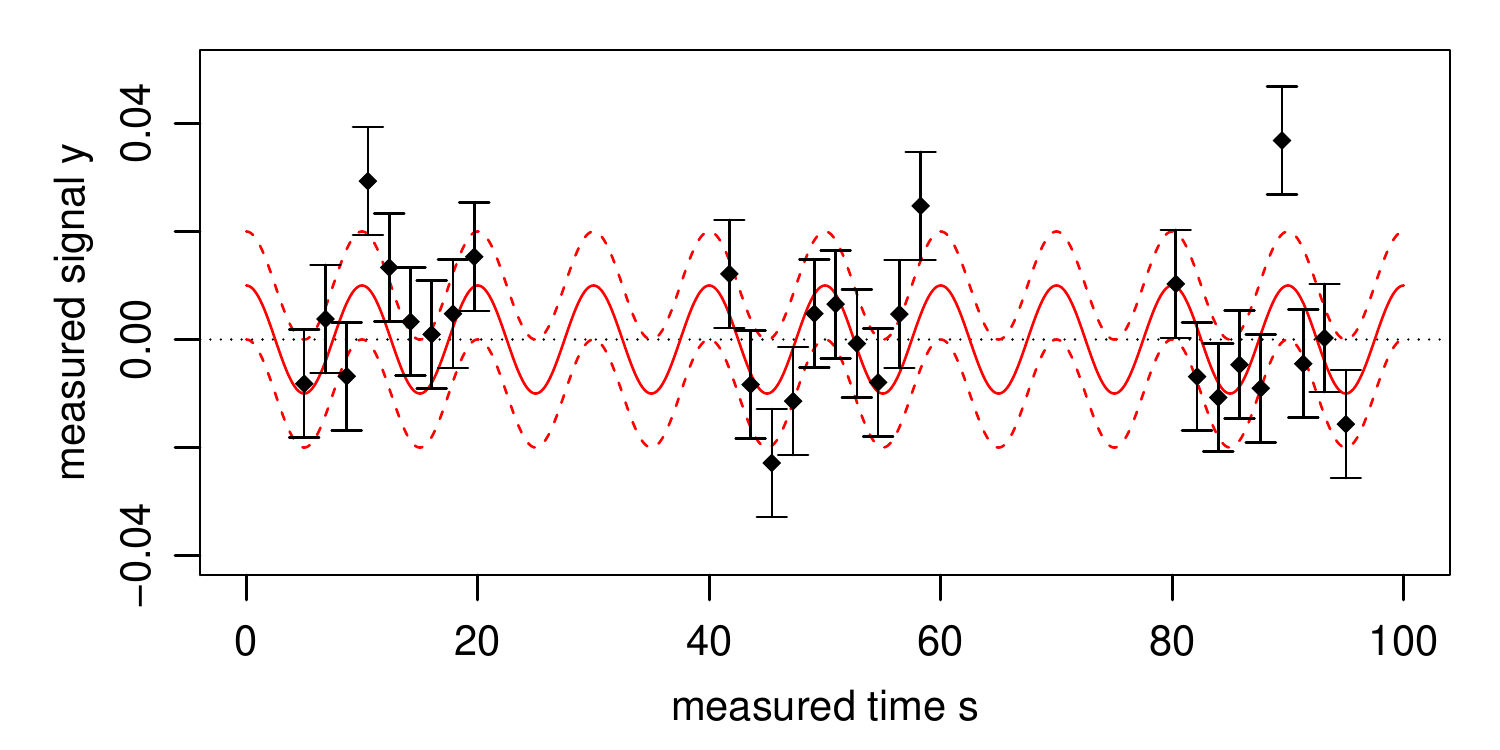}
\caption[]{Simulated sinusoidal data set. The solid red line is the true model; the dashed lines show the $\pm \sigma_{\rm true}$ variation about this. The black points show the data set drawn from this model along with their $\pm \sigma_{\rm true}$ error bars.
\label{fig:simrun06_lcurve}}
\end{center}
\end{figure}

In this section I demonstrate the method by applying it to data drawn from a known model. The true model, $z(t)$, is a sinusoid (equation~\ref{eqn:tsmod1sinusoid}) with amplitude $a=0.02$, frequency $\nu=0.1$, phase $\phi=0$, and background $b=0$. Fifty event times are drawn from a uniform random distribution between $t=5$ and $t=95$. There is no stochastic component. The measured signal, $y$, at each time is simulated by adding to $z$ Gaussian random noise with zero mean and standard deviation $\sigma_{\rm true}=0.01$. In order to simulate a typical astronomical light curve -- one with long gaps corresponding to day time -- events lying in the range $t=$\,20--40 and $t=$\,60--80 were removed, leaving 28 events. (We can consider $t$ as being in units of hours and the signal in units of magnitudes.) The true model and measured data are shown in Figure~\ref{fig:simrun06_lcurve}. (Numerous other experiments on other light curves have also been performed.)

I apply five models to the data, constructed by combining components in Table~\ref{tab:tsmodels}.
\begin{itemize}
\item Off+Sin+Stoch: the single frequency sinusoidal model with offset
  for TSMod1 (equation~\ref{eqn:tsmod1sinusoid}), and the Gaussian
  stochastic subcomponent for TSMod2 (equation~\ref{eqn:tsmod2gaussian}).
  The five adjustable parameters are $\theta = (a, \nu, \phi, b, \omega)$.
\item Sin+Stoch: as Off+Sin+Stoch, but with the offset $b$ fixed to zero (four adjustable parameters).
\item Sin: as Sin+Stoch, but with the stochastic component in the signal removed ($\omega \rightarrow 0$; see section~\ref{sect:bypass} for how this is achieved) (three adjustable parameters).
\item Off+Stoch: a simple offset (TSMod1) with a Gaussian stochastic subcomponent for TSMod2. This model has two adjustable parameters $\theta = (b, \omega)$.
\item OUprocess: the time series is modelled using the Ornstein--Uhlenbeck stochastic process, as described in appendix~\ref{sect:fullystochastic}. This has five adjustable parameters $\theta = (\tau, c, \mu[z_1], V[z_1], b)$
\end{itemize}


\begin{table}[!t]
\begin{center}
\caption{Log (base 10) LOO-CV likelihood of each model relative to that for the no-model
($\log L_{\rm LOO-CV} - \log L^{\rm NM}$), simulated sinusoidal light curve.
The last column gives the log likelihood for the no-model, $\log L^{\rm NM}$.
The first row is for models using the true value of the signal standard deviation, $\{\sigma_{y_j}\} = \sigma_{\rm true} = 0.01$.  The second row is for models using half the true value.
\label{tab:simrun0607kfold}}\vspace*{1em}
\begin{tabular}{lrrrrrr}
\toprule
$\{\sigma_{y_j}\}$ & OUprocess & Off+Stoch & Sin & Sin+Stoch & Off+Sin+Stoch & no-model \\
\midrule
0.01     & -0.74 & 0.19   &  2.63  & 2.04   & 2.40   & -21.95 \\
0.005   & 22.01 & 24.16 & 12.67 & 25.93 & 25.04 & -45.85 \\
\bottomrule
\end{tabular}
\end{center}
\end{table}

Model parameter priors and MCMC parameters are set following the criteria described in section~\ref{sect:modcomp}. The method is then applied -- as described in section~\ref{sect:application} -- to each of the five models, for two cases: (1) the
standard deviations in the measured signal, $\{\sigma_{y_j}\}$, which are our estimated uncertainties, are all set equal to the true value, $\sigma_{\rm true}=0.01$; (2) $\{\sigma_{y_j}\}=\sigma_{\rm true}/2=0.005$, i.e.\ the uncertainties are underestimated by a factor of two (the light curve itself is not changed). The resulting values of the LOO-CV log likelihoods are shown in Table~\ref{tab:simrun0607kfold}.

Looking at the first row in this table, we see that all three sinusoidal models are significantly favoured 
(difference greater than 1) over the no-model (which has a relative LOO-CV log likelihood of zero by construction), the OU process and the Off+Stoch model.
The true model, Sin, achieves the highest likelihood, although the likelihood is not significantly lower in the other two sinusoidal models. 
Inspection of the 1D posterior PDFs of the Sin model for the 28 partitions shows that the posterior peaks around the true
frequency and amplitude in most of the partitions, although it is relatively broad over amplitude. The phase posterior PDF peaks sharply near to 0 or 1 in about half the partitions, but in the rest is broader or at intermediate values. Inspection of the posterior PDFs over the two extra parameters in the Off+Sin+Stoch model -- the offset, $b$, and the stochastic component standard deviation, $\omega$ -- shows that the mean for the offset has an average across the partitions of about 0.0025. 
This just reflects the fact that this particular data set has a small positive mean signal (of 0.0019).
$b=0$ generally lies within 1 standard deviation of the mean of the posterior PDF.
The mean value of $\omega$ in this five parameter model is also not zero, but ranges between 0.004 and 0.007 across the 28 partitions.

Overall, we see a reasonably confident detection of the true model, by a factor of 100 in likelihood relative to the non-sinusoidal models, and this with a relatively sparse, non-uniform data set and low signal-to-noise ratio ($a/\sigma_{y_j} = 2$ for all $j$). Unsurprisingly it is not possible to distinguish between the different sinusoidal models. Although these provide some evidence for a non zero $b$ and $\omega$, it is not enough to formally favour Sin+Stoch or Off+Sin+Stoch over Sin.

In the second case (second row of Table~\ref{tab:simrun0607kfold}) the true model is now Sin+Stoch, because the supplied signal standard deviation is now half the true value: an extra stochastic term is needed to explain the missing variance. This reduces the likelihood of the no-model. The LOO-CV likelihoods of all models are now much larger relative to this.
The Sin model is poorer than the other models, because it too lacks the stochastic term needed to explain the missing variance.
In contrast, Off+Stoch has a far higher likelihood: It is more important to have the stochastic component than the sinusoidal one in order to explain these data. The most favoured model is the true one, Sin+Stoch. Its posterior PDF over $\omega$ is bell-shaped with a mean of 0.01 and a standard deviation of 0.002 in essentially all partitions. 
Given that the error bars supplied with the data were 0.05 and not zero, we might expect that a value of $\omega$ less than 0.01 would be sufficient to explain the variance on top of the sinusoid. A larger value is needed, however, because this data set just happens to have a larger actual variance than explained by $\sigma_{\rm true}$, as we also saw in the first case.

In this example we have seen that the models lacking the stochastic component have likelihoods which are very sensitive to the estimated signal standard deviations. As these are usually hard to estimate accurately, we should generally use a model with a stochastic component (TSMod2) with a free parameter in order to accommodate such missing variance. 
Comparing results from this with those from a model without such a component will help us establish whether the additional variance is required.

\section{Application to astronomical light curves}\label{sect:astroapp}

\subsection{Background and data}

I now apply the method to a set of (sub)stellar light curves. Each light curve shows the variation over time of the total light received (in the $I$ band) from a very low mass star or brown dwarf, objects collectively referred to as ultra cool dwarfs (UCDs). The variability of these sources has been the subject of numerous studies, because the light curves may 
reveal something about the processes operating in these objects' atmospheres (e.g.\ Morales-Calder\'on et al.\ 2006, Bailer-Jones 2008, Radigan et al.\ 2012). Variability could plausibly occur on timescales of hours to tens of hours due to the evolution of surface features, which might be star spots or dust clouds. If the opacity or brightness of these surface features differs from the rest of the (sub)stellar photosphere, a change in the coverage of the features with time would modulate the integrated light received by the observer (the stars are not spatially resolved). Another plausible cause of variability is the star's rotation when it has inhomogeneous (but otherwise stable) surface features. (The rotation periods of these objects have been measured to be between a few hours and a few days, e.g.\ Bailer-Jones 2004, Reiners \& Basri 2008.) In general, both mechanisms could generate observable variability.

\begin{figure*} 
\begin{center}
\includegraphics[width=0.95\textwidth, angle=0]{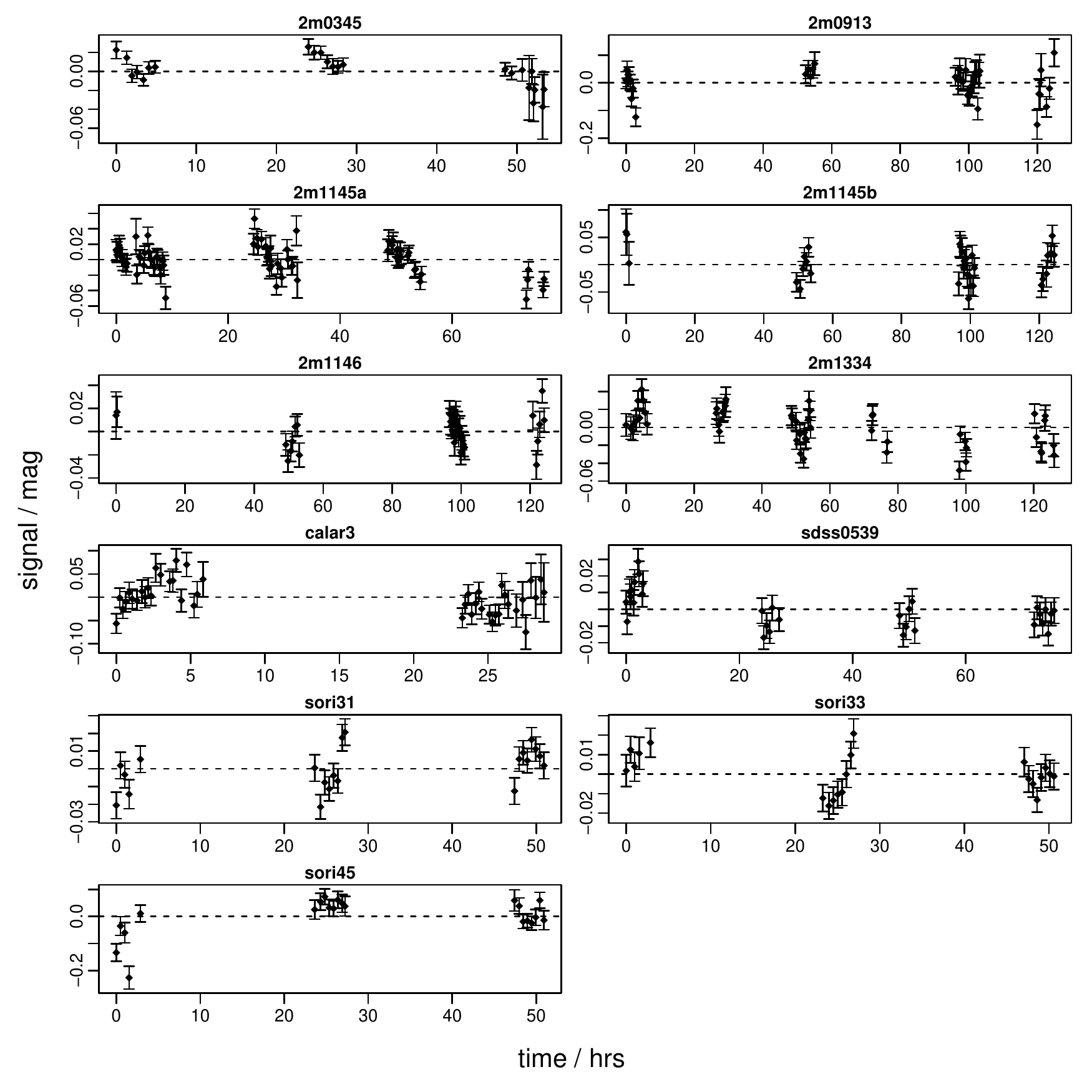}
\caption{Ultra cool dwarf light curves, showing the variation in brightness over time. Note that increases in brightness are {\em downwards} in the y axis (more negative magnitudes). 2m1145a and 2m1145b are light curves of the same object, but observed more than a year apart.
\label{fig:bdvar_lcurves}}
\end{center}
\end{figure*}

Here I use a set of 11 light curves previously obtained and analysed by Bailer-Jones \& Mundt (2001; hereafter BJM).\footnote{Tables 1 and 2 of BJM lists the properties of the objects. There are two light curves obtained at different times for the object 2M1145. The label 2m1145a is used in the present work to indicate the shorter duration light curve in Table 2 of BJM, i.e.\ the one with $t_{\rm max}$\,=\,76\,hours. 2m1145b labels the longer duration one.} 
The data are shown in Figure~\ref{fig:bdvar_lcurves}. In BJM, the light curves were subject to a simple orthodox hypothesis test using the $\chi^2$ statistic. The null hypothesis was that the light curve was constant, with fluctuations due only to heteroscedastic Gaussian noise, with zero mean and standard deviations estimated in the data reduction process. (This is  equivalent to the no-model in the present paper.) The p-value of the $\chi^2$ statistic was calculated, and if less than 0.01 the object was declared as being ``variable''.\footnote{BJM then go on to look for significant peaks in the CLEAN periodogram, and identify the variability as non-periodic if there is no significant peak in the periodogram. This suggests that the variability is not due to a simple rotational modulation, according to what was later called the ``masking hypothesis'': the rapid evolution of surface features ``erases'' the rotational modulation signature from the periodogram.} 
Of the 22 light curves analysed in BJM, 12 were declared as variable in this way, of which 11 are analysed here. (The other 11 light curves are no longer available unfortunately.)

Although this statistical test is widely used in this and other contexts, it is vulnerable to some of the standard -- and valid -- criticisms of orthodox hypothesis testing (see CBJ11 and references therein for further discussion). These are: the p-value is defined in terms of the probability of the $\chi^2$ being as large as or larger than the value observed, i.e\ it is defined in terms of unobserved and therefore irrelevant data; the p-value does not measure the probability of the hypothesis given the data, so does not answer the right question; a low p-value is used to reject the null hypothesis and therefore accept the (implicit) alternative, but without ever actually testing any alternative, even though the alternative may explain the data even less well. This final point is important, because all but the most trivial models generally give a very low probability for any particular data set, so a low p-value per se tells you little. What is important is the relative likelihoods of different models.  At best, a small p-value is just an indication that the null hypothesis may not be adequate to explain the data, but it is not a substitute for proper model assessment, i.e.\ model comparison. The onus is then on us to define alternative models and compare them in a suitable way, which is what I do here.
I use the same five models as were used in section~\ref{sect:simulations}

Although the measured data points have negligible timing uncertainties, they do have a finite duration (the integration time of the observations), either 5 or 8 minutes (fixed for a given light curve). This could be accommodated into the measurement model (section~\ref{sect:measmod}) by using a top-hat distribution instead of a Gaussian. I nonetheless approximate this as a delta function, for two reasons: (1) it accelerates considerably the likelihood calculations, because it allows us to replace the 2D likelihood integral (equation~\ref{eqn:eventlike}) with an analytic calculation (section~\ref{sect:errsmall}); (2) the method of calculating the likelihood of the OU process (derived in section~\ref{sect:OUprocessLike}) is only defined for this limit.

\subsection{Results: LOO-CV likelihood}

\begin{table}[!t]
\begin{center}
\caption{Log (base 10) LOO-CV likelihood of each model relative to that for the
  no-model for each light curve ($\log L_{\rm LOO-CV} - \log L^{\rm NM}$). The penultimate column gives the value of
the log likelihood for the no-model, $\log L^{\rm NM}$. The last column is the p-value for the hypothesis test from BJM.\label{tab:bdvarmainres}}\vspace*{1em}
\begin{tabular}{lrrrrrrr}
\toprule
light curve & OUprocess & Off+Stoch & Sin & Sin+Stoch & Off+Sin+Stoch & no-model & p-value\\
\midrule
2m0345 & 3.26 & 2.07 & 0.15 & 2.06 & 2.66 & -13.60 & 4e-4 \\  
2m0913 & 0.44 & 0.72 & 0.23 & 0.97 & 0.10 & -53.39 & 7e-4 \\  
2m1145a & 15.23 & 8.59 & 3.01 & 12.26 & 11.70 & -63.83 & $<$1e-9 \\  
2m1145b & -0.73 & 1.96 & 2.00 & 2.69 & 2.95 & -39.71 & 1e-3 \\  
2m1146 & 0.67 & 0.56 & -0.08 & 0.21 & 1.17 & -26.83 & 3e-3 \\ 
2m1334 & 14.95 & 12.82 & 4.06 & 16.86 & 16.12 & -65.88 & 1e-9 \\ 
sdss0539 & 5.50 & 1.99 & 4.93 & 4.48 & 4.67 & -19.62 & 3e-5 \\ 
calar3 & 3.60 & 1.43 & 5.65 & 5.11 & 4.28 & -28.06 & 6e-4 \\ 
sori31 & 2.04 & 2.12 & 1.02 & 2.59 & 1.90 & -11.16 & 4e-5 \\ 
sori33 & 1.49 & 0.66 & 2.14 & 1.85 & 2.12 & -8.39 & 2e-3 \\ 
sori45 & 6.70 & 4.32 & 5.08 & 6.23 & 6.32 & -29.93 & 5e-9 \\ 
\bottomrule
\end{tabular}
\end{center}
\end{table}

I follow the procedure outlined in section~\ref{sect:application} to define the priors and to sample the posterior with MCMC.  The results are summarized in Table~\ref{tab:bdvarmainres}.
A first glance over the table shows that for ten of the light curves, most of the models are significantly better than the no-model at explaining the data, often by a large amount. 

According to the $\chi^2$ test of BJM, all of these objects have a variability which is inconsistent with Gaussian noise on the scale of the error bars, so there should be a better model than the no-model (although it may not be among those I have tested.) We see from the Table that the no-model is not favoured for any light curve. However, for 2m0913, none of the models is {\em significantly} more likely than the no-model, so there is no reason to ``reject'' it. As the no-model is equivalent to the null hypothesis of BJM's $\chi^2$ test,  and this gave a p-value of $7e^{-4}$, this shows that the p-value is not a reliable metric for ``rejecting'' the null hypothesis.

On the other hand, in the three cases where the p-value is very low -- 2m1145a, 2m1334, sori45 -- the relative log likelihood for at least one model is high.
This suggests that a very low p-value {\em sometimes} correctly indicates that another model explains the data better, although this is of limited use as we do not know {\em how} low the p-value has to be. But at least it might motivate us to define and test other models. The converse is not true: a relatively high p-value does not indicate that the null hypothesis is the best fitting model. 

We turn now to identifying the best models. For all light curves, there is no significant difference between Off+Sin+Stoch and Sin+Stoch, which just means that the offset is not needed. That is not surprising, because the light curves have zero mean by construction. For 
eight of the light curves, the LOO-CV likelihood for Sin+Stoch is significantly larger than for Sin, implying there is a source of (Gaussian) stochastic variability which is not accounted for by the error bars in the data, $\{\sigma_{y_j}\}$.
This indicates either an additional source of variance (variability), or that 
the error bars have been underestimated.  (In only two of these cases -- 2m1145a and 2m1334 -- are the differences between Sin and Sin+Stoch very large.) 

Of course, there is no reason a priori to assume that a sinusoidal model is the appropriate one. In 9 of the 11 light curves, the sinusoidal models give a higher likelihood than the Off+Stoch model, and in the other two cases the value is not significantly lower. We can therefore state that for none of the 11 light curves is Off+Stoch significantly better than the sinusoidal models. But only with five or six light curves can we say that a sinusoidal model is {\em significantly} better than Off+Stoch. For the remaining light curves, the data (and priors!) do not discriminate sufficiently between the models, so neither can be ``rejected''.

Turning now to the OU process, we see that this is significantly better than all other models only for 2m1145a, but by a confident margin. In seven other cases the OU process is still better than the other models, or at least not significantly worse than the best model, so cannot be discounted as an explanation. In the remaining three cases -- 2m1145b, 2m1334, calar3 -- at least one other model is significantly better than the OU process.

The results for 2m1145a and 2m1145b are interesting, as these are light curves of the same object observed a year apart. At one time the OU process is the best explanation, at the other either a sinusoidal model or Off+Stoch.
Although it is plausible that the object shows different behaviour at different times, e.g.\ according to the degree of cloud coverage, we should not over-interpret this. We should also not forget that another, untested model could be better than any of these.

To summarize: Based just on the LOO-CV likelihood, I conclude that 10 of 11 light curves are explained much better by some model other than the no-model, by a factor of 100 or more in likelihood.
The exception is 2m0913, for which all models are equally plausible (likelihoods within a factor of ten).
Three light curves can be associated with one particular model:
2m1145a is best described by the OU process; 2m1334 and calar3 are best described by a sinusoidal model, the former requiring an additional stochastic component (Sin+Stoch), the latter could be either with or without it (Sin). 
This would seem to be consistent with a rotational modulation of the light curve (but see the next section).
For the remaining seven light curves, no single model emerges as the clear winner, although some models are significantly disfavoured. In three of these seven cases -- 2m0345, sdss0539,  sori45 -- both the OU process and a sinusoid model explain the data equally well (for 2m0345 and sori45 the sinusoidal model needs a stochastic component). For the remaining four light curves -- 2m1145b, 2m1146, sori31, sori33 --  the Off+Stoch model is at least as plausible as the other models. 
This model describes the data as having a larger Gaussian variance than is described by the error bars (with a possible constant offset to the light curve in addition). This could betray a variance intrinsic to the UCD, but it could equally well indicate that the error bars have been underestimated, something which is quite plausible given the multiple stages of the data reduction and approximations therein.

\subsection{Results: posterior PDFs}

To calculate the LOO-CV likelihood for a light curve with $J$ events, we had to sample from $J$ different posterior PDFs -- one per partition -- each given by equation~\ref{eqn:postPDF}. 
Here I examine the posteriors for the three light curves which could be associated with one particular model:
2m1334, calar3, and 2m1145a.

\begin{figure}
\begin{center}
\includegraphics[width=0.85\textwidth]{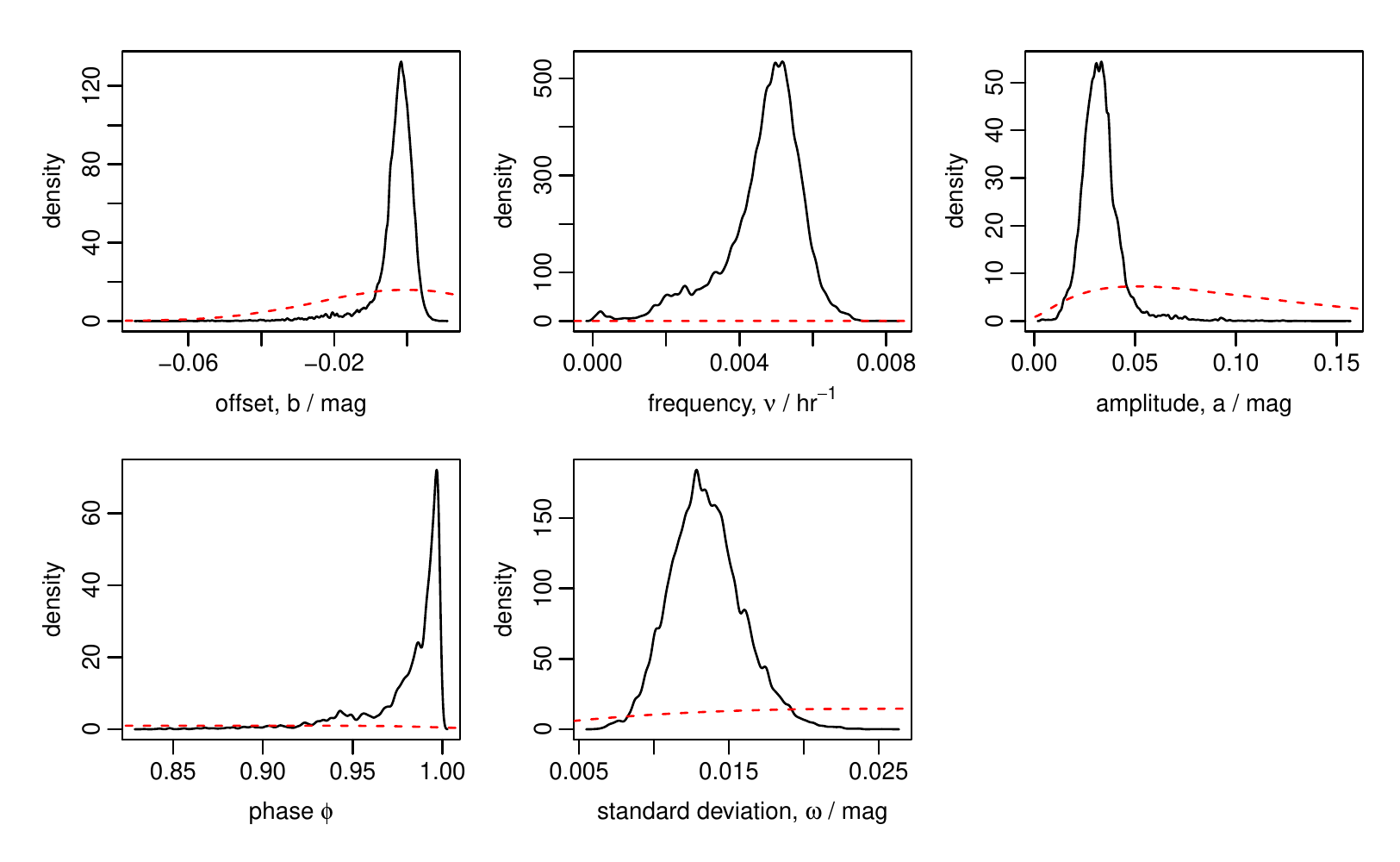}
\caption[]{Posterior PDF (black solid line) and prior PDF (red dashed line) over the five parameters in the Off+Sin+Stoch model of 2m1334, for one of the data partitions. The posterior PDFs for most of the other partitions look similar.
\label{fig:2m1334}}
\end{center}
\end{figure}

Figure~\ref{fig:2m1334} shows the 1D projections of the 5D posterior PDF for the Off+Sin+Stoch model on the 2m1334 light curve. The most probable model was Sin+Stoch, and the PDFs over the parameters this model has in common with Off+Sin+Stoch are similar.
In the first panel we see that the offset is consistent with being zero, as expected. 
Most of the probability for the frequency (second panel) lies between 0.004 and 0.006\,hr$^{-1}$, or periods of 170--250\,hr. This is considerably larger than the periods
$6.3\pm0.4$\,hr ($\nu=0.16$\,hr$^{-1}$) and $1.01\pm0.08$\,hr
identified by BJM (at a signal-to-noise ratio of 6 and 7 respectively).
170--250\,hr is also relatively long for a rotation period for a UCD and longer than the duration of the light curve. Thus the models to which this frequency range corresponds are in fact not periodic (no complete cycle) but just long term trends. A visual inspection of the light curve supports this. This could be intrinsic to the UCD or could be a slow drift of the zero point of the photometric calibration.

A more detailed study could overcome this by introducing an explicit trend model which is distinct from a periodic model. The prior PDF over frequency of the sinusoidal model would then be truncated at low values to ensure that such a model is truly periodic. This was done in CBJ11, where the evidence was calculated by averaging the likelihoods over a limited range of the period parameter. 

Examining further Figure~\ref{fig:2m1334}, we see strong evidence for a non-zero value of $\omega$ (the prior permits much smaller values), indicating that we need a stochastic component to explain variance on top of the (low frequency) sinusoidal component. Note finally that the posterior PDFs are far narrower than the corresponding prior PDFs (plotted as red dashed lines), which are essentially flat for three of the parameters (cf.\ Figure~\ref{fig:gammaPDF}).  This suggests that the priors are relatively uninformative, so the results are not very sensitive to their exact choice. 

\begin{figure}
\begin{center}
\includegraphics[width=0.85\textwidth]{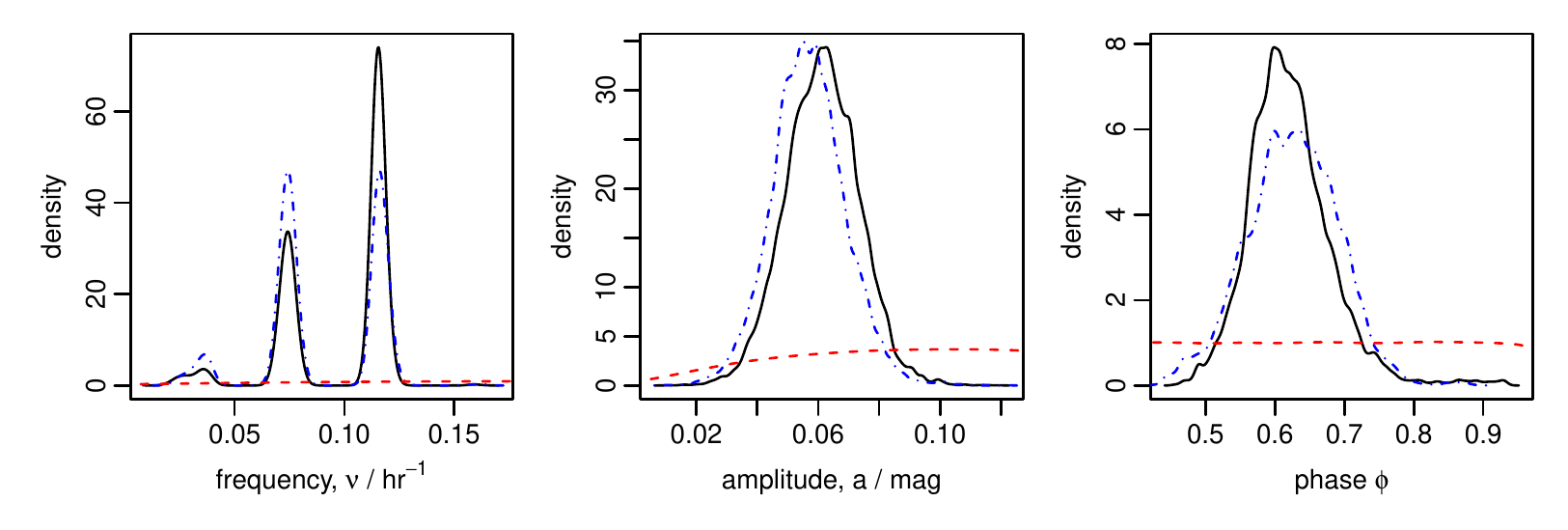}
\caption[]{Posterior PDF (black/solid line for one partition, blue/dot-dashed for another) and prior PDF (red dashed line) over the three parameters of the Sin model of calar3. The posterior PDFs for the other partitions are likewise very similar.
\label{fig:calar3}}
\end{center}
\end{figure}

The other light curve for which a sinusoidal model is significantly favoured is calar3. The posterior PDFs of the three parameters of the Sin model are shown in Fig.~\ref{fig:calar3} for two partitions. The PDFs are very similar for these and all other partitions. We see two distinct peaks in frequency, at around 0.075\,hr$^{-1}$ and 0.12\,hr$^{-1}$, or periods of 13.3\,hr and 8.3\,hr.  Both are plausible rotation periods. In comparison, BJM found peaks in the CLEAN periodogram at 14.0\,hr and 8.5\,hr, although at less than five times the noise they were described as ``barely significant''. These nonetheless agree with the periods found in the present analysis to within the uncertainties. As the integrated probability in these peaks in Figure~\ref{fig:calar3} is very large, then if Sin is the correct model (and not just the most probable of those tested) then these periods are significant.  The double-peaked posterior PDF means that there is evidence supporting models at both periods (or one is an alias), but Sin is still a single component model. In particular, the posterior PDF over amplitude in Fig.~\ref{fig:calar3} is the distribution over amplitudes for all models, i.e.\ for the entire frequency range. In order to determine the best fitting amplitudes for a model with two sinusoidal components we would need to calculate the posterior PDF for a six parameter model.

Of the 11 light curves explored here, BJM identified significant periods for 2m1146, 2m1334, sdss0539, and sori31. The present analysis suggests all of these {\em could} be explained by a periodic model, but only for 2m1334 is the periodic model significantly better than the others (although we just saw that this ``period'' is actually a trend). Furthermore, for 2m1146 only the Off+Sin+Stoch model is more significant than the no-model, and then only barely (LOO-CV log likelihood difference of just 1.17). So there is no strong evidence supporting periodicity in these four light curves.

\begin{figure}
\begin{center}
\includegraphics[width=0.85\textwidth]{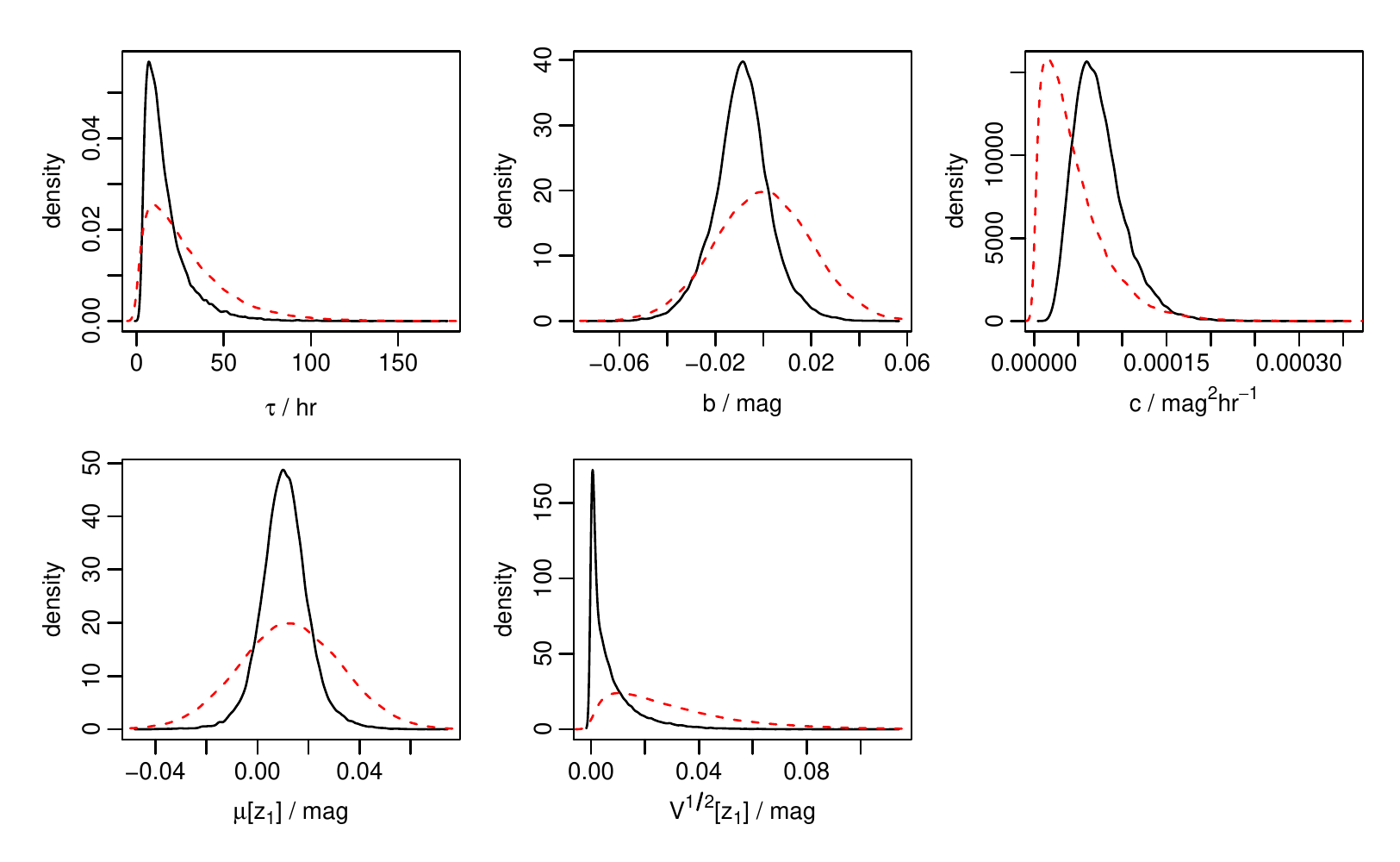}
\caption[]{Posterior PDF (black solid line) and prior PDF (red dashed line) over the five parameters in the Off+Sin+Stoch model of 2m1145a, for one of the partitions. The posterior PDFs for the other partitions look virtually identical.
\label{fig:2m1145}}
\end{center}
\end{figure}

The only other light curve with a clear ``winner'' model in terms of the LOO-CV likelihood is 2m1145a, for which the OU process was identified.
The PDFs over one partition are shown in Figure~\ref{fig:2m1145}.
Noticeable here is that the posterior PDFs are not much narrower than the prior PDFs. In Bayesian terms, the Occam factor (the ratio of volume of the parameter space occupied by the posterior to that occupied by the prior) is not much smaller than one, which means the data are not providing a strong discrimination over parameter solutions. This can be interpreted to mean that the model is quite flexible: a wide range of parameter settings are able to explain the data. This is perhaps not surprising given the nature of the OU process and the low signal-to-noise ratio of the data (the standard deviation of the signal, $\varsigma_y$, is only 1.7 times the mean error bar, $\overline{\sigma_{y_j}}$).
Alternatively, we can interpret this similarity between prior and posterior to mean that we used a comparatively informative (narrow) prior -- see next section.

\subsection{Results: sensitivity of the LOO-CV likelihood to the prior}

An important issue discussed in section~\ref{sect:CV} is the sensitivity of results to the priors adopted. I investigate this here by increasing the scale over which the prior PDFs extend. Specifically, I multiply by 2, then by 4, the standard deviation of Gaussian priors and the scale parameter of beta priors, then re-run the MCMC to recalculate the LOO-CV likelihood. The results of doing this for the sdss0539 light curve are shown in Table~\ref{tab:sdss0539kfoldpriorsens}. Although the likelihoods do change, they do not change by more than the factor of ten adopted here to indicate a significant difference. This pattern is generally seen with the other light curves also. However, there are a few cases in which 
we can get larger changes in the likelihood for apparently innocuous changes in the priors. This requires further study.

\begin{table}[!t]
\begin{center}
\caption{Log (base 10) LOO-CV likelihood of each model relative to that for the
  no-model ($\log L_{\rm LOO-CV} - \log L^{\rm NM}$), for the light curve sdss0539. Each line corresponds to different settings of the parameters of the prior PDFs. \label{tab:sdss0539kfoldpriorsens}}\vspace*{1em}
\begin{tabular}{lrrrrr}
\toprule
priors & OUprocess & Off+Stoch & Sin & Sin+Stoch & Off+Sin+Stoch  \\
\midrule
canonical & 5.50 & 1.99 & 4.93 & 4.48 & 4.67 \\
$\times$2 scale   & 5.14 & 2.02 & 4.51 & 4.11 & 4.13 \\
$\times$4 scale   & 4.60 & 1.98 & 5.08 & 4.51 & 4.83 \\
\bottomrule
\end{tabular}
\end{center}
\end{table}

\subsection{Results: Bayesian evidence}

I have introduced the LOO-CV likelihood as an alternative to the Bayesian evidence, on the basis that it is less sensitive to changes in the prior. I investigate this here.

As the no-model has no adjustable parameters, its evidence is equal to its likelihood and its LOO-CV likelihood.  I therefore use this again as a baseline against which to report the evidence. This relative evidence is reported in Table~\ref{tab:sdss0539evidencepriorsens} for sdss0539, for the same models and priors as shown in Table~\ref{tab:sdss0539kfoldpriorsens}. Comparing the first lines between these tables (canonical priors), we arrive at a similar conclusion based on the evidence as
based on the LOO-CV likelihood:  the OU process is the single best model; Off+Stoch and the no-model are significantly less likely. However, based on the evidence we would now say that the sinusoidal models are also significantly less likely.
Examining how the evidence changes with the priors, we see significant (more than unity) changes for all models, something we do not see for the LOO-CV likelihood in Table~\ref{tab:sdss0539kfoldpriorsens}. The evidence is indeed more sensitive to the priors in this case. We observe a similar behaviour for other light curves (although there are cases where doubling the scale of the priors changes the evidence by less than a factor of ten).
Prior sensitivity nonetheless remains an issue for LOO-CV likelihood, and this should always be investigated in any practical application.

\begin{table}[!t]
\begin{center}
\caption{Log (base 10) evidence of each model relative to that for the
  no-model, $\log E - \log L^{\rm NM}$, for the light curve sdss0539.
Each line corresponds to different settings of the parameters of the prior PDFs. Cf.\ Table~\ref{tab:sdss0539kfoldpriorsens}.
\label{tab:sdss0539evidencepriorsens}}\vspace*{1em}
\begin{tabular}{lrrrrr}
\toprule
priors & OUprocess & Off+Stoch & Sin & Sin+Stoch & Off+Sin+Stoch  \\
\midrule
canonical &  5.38  &  1.07    &  3.24 &  2.31  &  2.03  \\
$\times$2 scale   &  4.72  &  0.30    &  2.42 &  0.65  & -0.90  \\
$\times$4 scale   &  3.97  &  -0.53  &  1.81 &  -0.87 & -2.63  \\
\bottomrule
\end{tabular}
\end{center}
\end{table}

\section{Summary and conclusions}\label{sect:summary}

This article has introduced three ideas
\begin{enumerate}
\item a fully probabilistic method for modelling time series with arbitrary temporal spacing. It can accommodate any kind of measurement model (error bars) on both the time and signal variables, as well as any functional model for the signal--time dependence and the stochastic variations in both of these. In contrast to many other time series modelling methods, it can model a stochastic variation in the time axis too.
It can in fact be used to model any 2D data set, and not just temporal data: with a linear model it offers an alternative solution to the total least squares solution for data sets with errors in both variables, for example.
\item a cross-validation alternative to the Bayesian evidence, which is based on the posterior-averaged likelihood (combined over partitions of the data) as opposed to the prior-averaged likelihood. In theory this is less sensitive to the prior parameter PDFs than the evidence, something confirmed by the initial experiments reported here. Experiments on simulated data suggest that this metric is an effective means of model comparison. Its main drawback in comparison to the evidence is that it takes longer to calculate.
\item the use of the Ornstein--Uhlenbeck process in a Bayesian time series model, i.e.\ one in which we sample rather than maximize the posterior. (Theoretical results similar to the event likelihood and the recurrence relation for the posterior PDF -- derived in appendix~\ref{sect:fullystochastic} -- have been published elsewhere.)
\end{enumerate}
The main purpose of this article was to give a detailed theoretical exposition of the model.
A more comprehensive application to time series analysis problems will be published elsewhere. In the present work I have demonstrated the method using simulated data, and through an analysis of 11 brown dwarf light curves. 
The main conclusions of this study are as follows
\begin{itemize}
\item 10 of 11 light curves are explained significantly better by one of the models tested than by the no-model, the ``null hypothesis'' that the variability is just due to Gaussian fluctuations with standard deviation given by the error bars about the mean of the data. ``Significantly better'' here means the LOO-CV likelihood is at least 100 times larger, something we might interpret as a 99\% confidence level.  
For comparison, in BJM all 11 light curves were flagged as variable at a (different) 99\% confidence level based on an orthodox $\chi^2$ test of the same null hypothesis. The Bayesian model comparison performed here has a sounder theoretical basis, gives us more confidence in the results, and supplies more information.
The two methods disagree on 2m0913, which is adequately explained by the no-model here. 
\item three light curves are described significantly better by one model than any of the others: 2m1145a by the OU process; 
2m1334 by a sinusoid with an additional Gaussian stochastic component; calar3 by a sinusoid either with or without an additional stochastic component. However, the probable periods for 2m1334 are longer than the duration of the time series, so this is best interpreted as a long-term trend rather than a periodic variation. For calar3 we see two distinct and significant peaks in the posterior PDF over the frequency, at 0.12\,hr$^{-1}$ (period=8.3\,hr) and 0.075\,hr$^{-1}$ (period=13.3\,hr). Both had been identified in earlier work, but as barely statistically significant.
\item the other 8 light curves can be described by more than one model, either the OU process, or a constant with stochastic component or a sinusoid with stochastic component. 
\end{itemize}
It must be remembered that we can only comment on models we have explicitly tested: it remains possible that other plausible models exist which could explain the data better. Future work with this method will focus on its practical application to scientific problems, the inclusion of more time series models, as well as further testing of the LOO-CV sensitivity to the prior PDFs.

\begin{acknowledgements}
I would like to thank Rene Andrae and the referee, Jeffrey Scargle, for useful comments.
\end{acknowledgements}

\appendix

\section{Fully stochastic time series processes}\label{sect:fullystochastic}

The signal component of the time series model is the PDF $P(z_j | t_j, \theta, M)$. 
For a physical process which has a well-defined, time-variable signal on top of which there is some randomness, 
section~\ref{sect:tsmodel} shows a convenient way of expressing this as two independent subcomponents: one which describes the time-dependence of the mean of the PDF and the other which describes the PDF itself and its time-independent parameters, e.g.\ its variance.

A fully stochastic process, in contrast, is one in which all of the parameters of the PDF can have a time dependence. Given a functional form for this time dependence, we can in principle just introduce this into $\theta$ and calculate the likelihood as before. 
A simple fully stochastic process is one with a constant mean and variance, a white noise process. This is achieved by setting TSMod1
to a uniform model ($\eta = b$ in equation~\ref{eqn:tsmod1sinusoid}) with TSMod2 a Gaussian.

However, incorporating a stochastic process which has memory, such as a Markov process, is more complicated. 
Here I show how to introduce a particular but widely used stochastic process, the Ornstein--Uhlenbeck process.

\subsection{The Ornstein--Uhlenbeck process}\label{sect:OUprocess}

The Ornstein--Uhlenbeck (OU) process (Uhlenbeck \& Ornstein 1930) is a stochastic process which describes the evolution of a scalar random variable, $z$. The equation of motion (Langevin equation) for $t>0$ can be written 
\begin{equation}
\label{eqn:OUdiffeqn}
dz(t) \,=\, - \frac{1}{\tau}z(t)dt + c^{1/2}{\cal N}(t ; 0, dt)
\end{equation}
where $\tau$ and $c$ are positive constants, the {\em relaxation time} (dimension $t$) and the {\em diffusion constant} (dimension $z^2 t^{-1}$) respectively, $dt$ is an infinitesimally short time interval, ${\cal N}(t ; 0, dt)$ is a Gaussian random variable with zero mean and variance $dt$, and $dz(t) = z(t+dt) - z(t)$. 
The OU process is the continuous-time analogue of the discrete-time AR(1) (autoregressive) process, and is sometimes referred to as the CAR(1) process. In the context of Brownian motion, $z(t)$ describes the velocity of the particle. 
There are alternative, equivalent forms of this equation of motion. 
For more details see Gillespie (1996). 

The OU process is stationary, Gaussian and Markov.\footnote{Put loosely: {\em Stationary} means that the joint PDF of a set of events from the process is invariant under translations in time; {\em Markov} means that the present value of the process depends only on the value at one previous time step;
{\em Gaussian} means that the joint PDF of any set of points is a multivariate Gaussian, in particular the PDF of a single point is Gaussian.} 
The PDF of $z(t)$ is Gaussian with mean and variance given by
\begin{subequations}
\label{eqn:OUprocess}
\begin{alignat}{2}
\mu_z \,&=\, z_0 \upsilon \label{eqn:OUprocess_mean} \\
V_z \,&=\, \frac{c\tau}{2}(1 - \upsilon^2 ) 
\end{alignat}
\end{subequations}
respectively, for any $t>t_0$, where $z_0 = z(t\!=\!t_0)$ and
\begin{equation}
\upsilon \,=\, e^{-(t-t_0)/\tau} \ .
\end{equation}
Given the initial condition $z_0$ at $t_0$, we know the PDF of the process at any subsequent time. The relaxation time, $\tau$, determines the time scale over which the mean and variance change. 
The diffusion constant determines the amplitude of the variance.
The OU process $z(t)$ is a mean-reverting process: for $t-t_0 \gg \tau$ the mean tends towards zero and the variance asymptotes to $c\tau/2$ (for finite $\tau$).
From this we can derive an update equation to give the value of the process at time $t$,
\begin{equation}
z(t) \,=\, z_0\upsilon\,+\, n_1 \sqrt{V_z}
\label{eqn:OUupdates}
\end{equation}
where $n_1$ is unit random Gaussian variable (Gillespie 1996). This is just the sum of the mean and a random number drawn from a zero-mean Gaussian with the variance at time $t$.
For a given sequence of time steps, $(t_0, t_1, \ldots)$, we can use this to generate an OU process.
Because the time series is stochastic and must be calculated at discrete steps, then even for a fixed random number seed, the generated time series depends on the actual sequence of steps.

The reader may be more familiar with the Wiener process. This can be considered a special case of the OU process in which $\tau \to \infty$ (Gillespie 1996b), in which case $\upsilon \to 1$. The update equation becomes $z(t) = z_0 + n_1 \sqrt{V_z}$, where now $V_z=c(t-t_0)$.

\subsection{Likelihood of the Ornstein--Uhlenbeck process}\label{sect:OUprocessLike}

A Markov process is one in which we can specify the PDF of the state variable, $z_j$ at time $t_j$, using
$P(z_j | t_j, z_{j-1}, t_{j-1}, \theta, M)$, i.e.\ there is a dependence on the previous state variable, $z_{j-1}$. 
For the OU process, this PDF is a Gaussian with mean and variance given by equation~\ref{eqn:OUprocess}.
Clearly, the nearer $t_{j-1}$ is to $t_j$ the better a measurement of $z_{j-1}$ will constrain $z_j$.

We could therefore write the signal component of the time series model (see equation~\ref{eqn:tsmod}) as
\begin{alignat}{2}
P(z_j | t_j) \,&=\, \int_{t_{j-1}, z_{j-1}} P(z_j | t_j, z_{j-1}, t_{j-1}) P(z_{j-1}, t_{j-1} | t_j) \, dt_{j-1} dz_{j-1} \nonumber \\
                \,&=\, \int_{t_{j-1}, z_{j-1}} P(z_j | t_j, z_{j-1}, t_{j-1}) P(z_{j-1} | t_{j-1}) P(t_{j-1}) \, dt_{j-1} dz_{j-1}
\label{eqn:MarkovLike}
\end{alignat}
where conditional independence has been applied in the second line to remove the $t_j$ dependence from the second two terms.
Note that everything is implicitly conditioned on $M$ and its parameters $\theta$, but these have been omitted for brevity. 
The first term under the integral is the PDF for the Markov process we aimed to introduce. The second term is also a PDF for the Markov process but referred to the previous event. We could replace that with another 2D integral over $(t_{j-2}, z_{j-2})$
of exactly the same form as equation~\ref{eqn:MarkovLike}. We could then continue recursively to achieve a chain of nested 2D integrals going back to the beginning of the time series, and use that in our likelihood calculation. Although this is a plausible and general solution for a Markov process, it is not very appealing.

Fortunately a significant simplification is possible. Let us first neglect the time uncertainties. In that case the event likelihood (equation~\ref{eqn:eventlike}) becomes
\begin{equation}
P(D_j | \sigma_j, \theta, M) \,=\, \int_{z_j} P(y_j | z_j, \sigma_{y_j}) P(z_j | t_j, \theta, M) P(t_j | \theta, M) \, dz_j
\label{eqn:eventlike6}
\end{equation}
with $t_j\!=\!s_j$. $P(y_j | z_j, \sigma_{y_j})$ is the signal part of the measurement model (the time part has dropped out).
If this is Gaussian in $y_j - z_j$  (cf.\ equation~\ref{eqn:measmod}) and $P(z_j | t_j, \theta, M)$ is Gaussian in $z_j$, then equation~\ref{eqn:eventlike6} is a convolution of two Gaussians, which is another Gaussian, multiplied by $P(t_j | \theta, M)$ (which is independent of $z_j$). A general result is that if $f$ is a Gaussian with mean $\mu_f$ and variance $V_f$, and $g$ is a Gaussian with mean $\mu_g$ and variance $V_g$ then
\begin{equation}
\int_{-\infty}^{+\infty} f(y - z)g(z) dz = f \otimes g
\label{eqn:convolution}
\end{equation}
is a Gaussian with mean $\mu_f + \mu_g$ and variance $V_f + V_g$. For the Gaussian measurement model, $f$ is Gaussian in the argument $y_j-z_j$ with $\mu_f = 0$ and $V_f = \sigma_{y_j}^2$. $g$ is then the time series model.

We now turn specifically to the OU process in order to determine its time series model, $P(z_j | t_j, \theta, M)$. 
We can derive this from equation~\ref{eqn:OUupdates}.
With the state variable now written as $z_j$ rather than $z(t)$, this update equation is
\begin{equation}
z_j \,=\, z_{j-1}\upsilon \,+\, n_1\sqrt{V_z}
\label{eqn:OUupdates2}
\end{equation}
with
\begin{subequations}
\label{eqn:OUupdates2_aux}
\begin{alignat}{2}
\upsilon \,&=\, e^{-(t_j - t_{j-1})/\tau}\\
V_z \,&=\, \frac{c\tau}{2}(1 - \upsilon^2) \ .
\end{alignat}
\end{subequations}
$z_j$ has a Gaussian distribution (by definition of the OU process) with
mean and variance\footnote{These we calculate explicitly from equation~\ref{eqn:OUupdates2}. The variance is just the sum of the variances of the two terms in that equation. Recall that in general $V(fg) = f^2V(g) + g^2V(f)$, and that $V(\upsilon)=0$.} 
\begin{subequations}
\label{eqn:OUprior}
\begin{alignat}{2}
\mu[z_j] \,&=\, \mu[z_{j-1}]\upsilon \label{eqn:OUprior_mean}\\
V[z_j] \,&=\, V[z_{j-1}]\upsilon^2 + V_z \label{eqn:OUprior_var}
\end{alignat}
\end{subequations}
respectively (see also Berliner 1996).  
Specifically, $P(z_j | t_j, \theta, M)$ is a Gaussian with this mean and variance, which are specified by the parameters $\theta = (\mu[z_{j-1}], V[z_{j-1}], \upsilon, \tau, c)$.\footnote{If we instead had an actual value of $z_{j-1}$, then 
$P(z_j | t_j, \theta, M)$ would be Gaussian with mean $z_{j-1}\upsilon$, variance $V_z$, and $\theta = (z_{j-1}, \upsilon, \tau, c)$.}
We will look in a moment at how we estimate $\mu[z_{j-1}]$ and $V[z_{j-1}]$.

We can now write the likelihood, the result of the Gaussian convolution, equations~\ref{eqn:eventlike6} and~\ref{eqn:convolution}, as
\begin{alignat}{2}
P(D_j | \sigma_j, \theta, M)  \,&=\, P(t_j | \theta, M) \int_{z_j} P(y_j | z_j, \sigma_{y_j}) P(z_j | t_j, \theta, M) \, dz_j \nonumber \\
                           \,&=\, P(t_j | \theta, M) \, \frac{1}{\sqrt{2\pi V[y_j]}} \, \exp \left(\frac{-(y_j - \mu[y_j])^2}{2V[y_j]}\right)
\label{eqn:eventlike7}
\end{alignat}
where the mean and variance of this Gaussian are
\begin{subequations}
\label{eqn:OUlikelihood}
\begin{alignat}{3}
&\mu[y_j] \,&&=\, 0                    &&+ \mu[z_j] \\
&V[y_j]      \,&&=\, \sigma_{y_j}^2 &&+ V[z_j]     
\end{alignat}
\end{subequations}
respectively. Recall that $P(t_j | \theta, M)$ is just the time component of the time series model with $t_j\!=\!s_j$. Normally we will use a uniform model (equation~\ref{eqn:tsmod3uniform}), so this is just a constant.

To estimate $\mu[z_{j-1}]$ and $V[z_{j-1}]$ we make use of the data, $y_{j-1}$.
For an event $t_j$,  $P(z_j |  t_j, \theta, M)$ -- which has mean and variance given by equation~\ref{eqn:OUprior} -- is our estimate of the PDF of the state variable at $t_j$ prior to taking into account the measurement $y_j$. It is therefore the appropriate thing to use to calculate the likelihood of $y_j$, as we have done in equation~\ref{eqn:eventlike7}. But in parallel to this we want to use $y_j$ to improve our estimate of $z_j$, i.e.\ we want to calculate the posterior PDF of $z_j$. This is given by Bayes' theorem
\begin{equation}
P(z_j | y_j,  t_j) \propto P(y_j | z_j, t_j) P(z_j | t_j)
\end{equation}
(ignoring the normalization constant $1/P(y_j | t_j)$, and omitting a lot of dependencies). These two terms are again the measurement model (so the dependence on $t_j$ drops out) and the time series model,  both of which are Gaussian in $z_j$. Thus the posterior PDF over $z_j$ is also a Gaussian with mean and variance\footnote{
The product of two Gaussians $f$ and $g$
with means $\mu_f$ and $\mu_g$ and variances $V_f$ and $V_g$ is another Gaussian with
\begin{equation}
{\rm mean} \hspace*{1em} \frac{\mu_f V_g + \mu_g V_f}{V_f + V_g}
\hspace*{1em} {\rm and~~variance} \hspace*{1em}
\frac{V_f V_g}{V_f + V_g} \ .
\end{equation} }
\begin{subequations}
\label{eqn:OUposterior}
\begin{alignat}{1}
\mu^{\prime}[z_j] \,&=\, \frac{ y_jV[z_j] + \mu[z_j]\sigma_{y_j}^2 }{ V[z_j] + \sigma_{y_j}^2 } \\
V^{\prime}[z_j] \,&=\, \frac{ V[z_j]\sigma_{y_j}^2 }{ V[z_j] + \sigma_{y_j}^2 } 
\end{alignat}
\end{subequations}
respectively, where the prime symbol is used to distinguish these posterior moments from the prior ones in equation~\ref{eqn:OUprior}. 
It is these quantities which we then use at the {\em next} event as the estimates of the mean and variance of the state variable. Thus, at iteration (event) $j$, when we calculate equation~\ref{eqn:OUlikelihood} and hence the likelihood, we use $\mu^{\prime}[z_{j-1}]$ and $V^{\prime}[z_{j-1}]$ as our estimates of 
$\mu[z_{j-1}]$ and $V[z_{j-1}]$. This is how we introduce a dependence on the previous measurement (the Markov property). We then calculate the mean and variance of the posterior for $z_j$ using equation~\ref{eqn:OUposterior}, and will then use these in the next iteration. Thus we have a recurrence relation for the posterior PDF of $z_j$, at each iteration siphoning off the relevant quantities in order to calculate the event likelihood.

To initialize the process we must specify initial values $\mu[z_1]$ and $V[z_1]$. We use 
these in equation~\ref{eqn:OUlikelihood} to calculate $\mu[y_1]$ and $V[y_1]$ and hence the likelihood for the first
event, $y_1$, from equation~\ref{eqn:eventlike7}. We then calculate the posterior moments using equation~\ref{eqn:OUposterior}. For the next event, $j=2$, these posterior moments are assigned to $\mu[z_{j-1}]$ and $V[z_{j-1}]$ in equation~\ref{eqn:OUprior} and the likelihood calculated. The procedure is iterated through all the events.

\begin{figure}
\begin{center}
\includegraphics[width=0.6\textwidth]{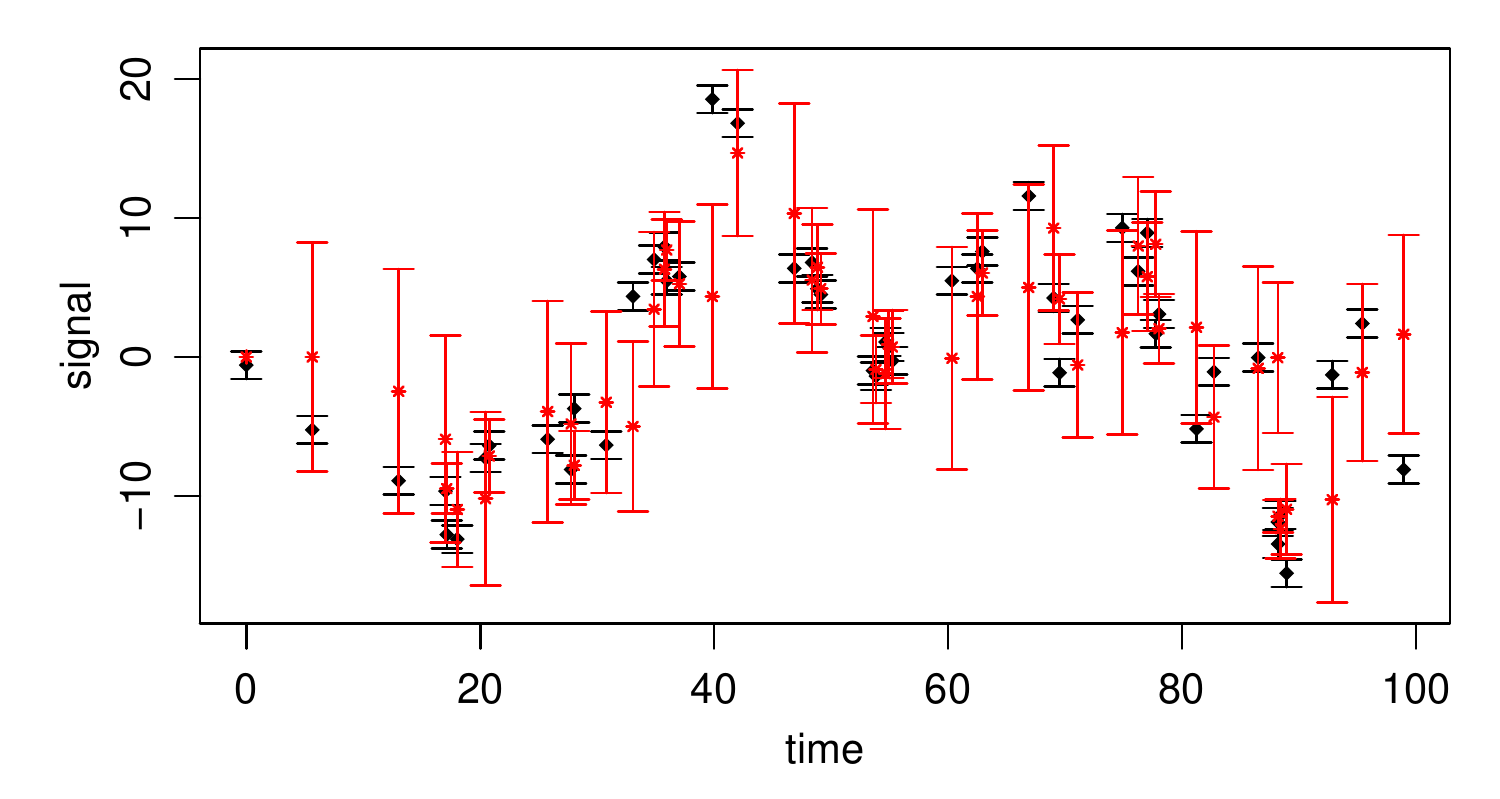}
\caption[]{Example simulated OU process. The black points (with a uniform random time distribution) have been simulated from an OU process with parameters $\tau=10$, $c=20$ and initial conditions $t_1=0$, $z_1=0$, to which Gaussian measurement noise with mean zero and unit standard deviation has been added (as indicated by the black error bars). This red points show the predictions of this process using the OU process time series model with parameters $\tau=10$, $c=20$, $\mu[z_1]=0$, $V[z_1]=0$. The prediction for each event is a Gaussian in 
the signal with mean and variance given equation~\ref{eqn:OUprior}.
\label{fig:lcurve_OUprocess}}
\end{center}
\end{figure}

The model prediction of the OU process is a Gaussian distribution at each event (at time $t_j$) with mean and variance given by equation~\ref{eqn:OUprior}.  Unlike the memoryless time series models, the OU process requires the measurement of the one previous event in addition to the model parameters in order to predict the next event (this is the Markov property). The relevant model prediction of event $j$ is therefore given by the prior distribution of equation~\ref{eqn:OUprior} -- which has not yet looked at $y_j$ -- and not by the posterior distribution of equation~\ref{eqn:OUposterior}, which has.

The parameters of the process are $\theta = (\mu[z_1], V[z_1], \tau, c)$ (and implicitly the initial time, $t_1$). 
Figure~\ref{fig:lcurve_OUprocess} shows an example of a simulated OU process and the model predictions thereof. 

As it stands, the long-term mean of this OU process is zero. We can introduce this long-term mean as an additional parameter, $b$, of the model. Equation~\ref{eqn:OUprior_mean} then becomes
\begin{equation}
\label{eqn:OUprior_mean_mod}
\mu[z_j] \,=\, \mu[z_{j-1}]\upsilon + b(1-\upsilon) \ .
\end{equation}
The variance is unchanged. (See also Brockwell \& Davis 2002, section 10.4.) Note that this corresponds to solving a different differential equation, namely one in which we have the additional term $(b/\tau)dt$ on the right-hand-side of equation~\ref{eqn:OUdiffeqn}.
(Note that $b$ is the {\em long-term} mean of the process rather than the mean of the data.)

Now that we can calculate the likelihood, we can calculate the evidence or sample the posterior PDF.
By partitioning the data set we can also use posterior sampling to evaluate the cross-validation likelihood, as described in section~\ref{sect:CV}. Note that whatever partitioning we do, when it comes to calculating the partition likelihood  for data $D_k$ we must still use all of the data to predict the full sequence of events.
That is, for a given $\theta_n$, we predict the entire sequence of $J$ events using all the data, but only make use of those event likelihoods which are appropriate. Specifically, to calculate the posterior to use in MCMC sampling (equation~\ref{eqn:postPDF}) we just select the likelihood for the events in $D_{-k}$, and to calculate the likelihoods in equation~\ref{eqn:partlike} (or~\ref{eqn:partlike2}) we just use the events in $D_k$. 
The OU process depends not only on the model parameters but also on the state at the previous time step, so we should not be changing these time steps by removing events when predicting the sequence.

\subsection{Literature note}

I am not aware of an explicit derivation in the literature either of the above posterior recurrence relation (although Berliner 1996 outlines the same thing)
or of the event likelihood for the OU process. Kelly et al. (2009) write down similar equations for the latter (their equations 6--12), but in a significantly rearranged form. 
Kelly et al.\ also assume a specific initial value (zero) for the initial state variable, $z_0$ (their equation 8), whereas I give this a distribution. 
(In my formulation we can achieve a specific initial value by setting $V[z_1]=0$.)
My expression for the variance (equation~\ref{eqn:OUprior_var}) therefore has an additional term compared to theirs (their equation A5, which is $V_z$ in my notation), because they are conditioning on an fixed value of the process at the previous step whereas I assume this itself has a variance, $V[z_{j-1}]$. (See also section 10.4 of Brockwell \& Davis 2002.) 
Closely related formulations of this process -- but not the likelihood calculation -- are given in Jones (1986; sections 4 and 5) and Kozlowski et al.\ (2010; appendix).

\section{Simplifying the event likelihood integration}\label{sect:simplifications}
 
The calculation of the likelihood for a single event in principle requires a 2D integration (equation~\ref{eqn:eventlike}).
This can, however, be reduced to a 1D integration or even just a function evaluation under certain circumstances.

\subsection{Dropping the stochastic signal component of the time series model (TSMod2 bypass)}\label{sect:bypass}

If TSMod2 is a Gaussian (equation~\ref{eqn:tsmod2gaussian})
in which $\omega$ is very small compared to the scale of signal variations, then the only contribution to the event likelihood is at the prediction of the signal by TSMod1. For given $\theta_1$, the magnitude of the event likelihood is then dictated only by the measurement model, i.e.\ how close the measured $y_j$ is to the prediction $z_j$. 
In this limit $\omega \rightarrow 0$, the signal part of the time series model (equation~\ref{eqn:tsmod}) becomes
$P(z_j | t_j, \theta_1, \theta_2, M) = \delta(z_j - \eta[t_j ; \theta_1])$. 
This gives us a purely deterministic signal in the time series model; we ``bypass'' TSMod2. The event likelihood integration (equation~\ref{eqn:eventlike}) then becomes a 1D integration\footnote{As we now have no stochastic element in either $t$ or $z$, the reader may wonder why this integral is over $t$ rather than $z$, i.e.\ why there is an asymmetry. The point is that we need to integrate along the path of the (deterministic) function $z_j=\eta[t_j ; \theta_1]$. As this only requires one parameter, we only have a one-dimensional integral. Whether we parametrize this with $t_j$ or $z_j$ is unimportant, but having written the function as $z_j=\eta[t_j ; \theta_1]$ rather than $t_j=\eta^\prime[z_j ; \theta_1]$, $t_j$ is the more natural choice.}
\begin{equation}
P(D_j | \sigma_j, \theta, M) \,=\, \int_{t_j} \underbrace{P(D_j | t_j, z_j=\eta[t_j ; \theta_1], \sigma_j)}_\text{Measurement model} \ \underbrace{ P(t_j | \theta_3, M)}_\text{Time series model} \ dt_j \ \ .
\label{eqn:eventlike2}
\end{equation}

\subsection{Small uncertainties on the measured times}\label{sect:errsmall}

If the uncertainty on the measured time, $\sigma_{s_j}$, is very small compared to the time scale over which the time series model varies, then the integral over $t_j$ in the event likelihood will have a significant contribution only for times $t_j$ close $s_j$. This must hold for any sensible measurement model or definition of uncertainties.
The time part of the measurement model can then be approximated by the delta function 
$\delta(t_j - s_j)$, and the integration over $t_j$ is just unity.
If the signal part of the measurement model is Gaussian, the event likelihood equation (equation~\ref{eqn:eventlike}) becomes
\begin{equation}
P(D_j | \sigma_{y_j}, \theta, M) \,=\, \int_{z_j} \underbrace{ \frac{1}{\sqrt{2\pi}\sigma_{y_j}} \, e^{-(y_j - z_j)^2/2\sigma_{y_j}^2} }_\text{Measurement model} \ \underbrace{ \vphantom{\frac{1}{\sqrt{2\pi}\sigma_{y_j}}} P(t_j=s_j, z_j | \theta, M) }_\text{Time series model} \ dz_j \ \ .
\label{eqn:eventlike3}
\end{equation}
When TSMod2 is the Gaussian model this becomes
\begin{equation}
P(D_j | \sigma_{y_j}, \theta, M) \,=\, \int_{z_j} \underbrace{ \frac{1}{\sqrt{2\pi}\sigma_{y_j}} \, e^{-(y_j - z_j)^2/2\sigma_{y_j}^2} }_\text{Measurement model} \ \underbrace{  \vphantom{\frac{1}{\sqrt{2\pi}\sigma_{y_j}}} \frac{1}{\sqrt{2\pi}\omega} \, e^{-(z_j - \eta[s_j ; \theta_1])^2/2\omega^2} \ P(s_j | \theta_3, M) }_\text{Time series model} \ dz_j \ .
\end{equation}
This can be written as
\begin{equation}
P(D_j | \sigma_{y_j}, \theta, M) \,=\,  P(s_j | \theta_3, M) \int_{z_j} f(y_j-z_j) g(z_j) dz_j \ .
\end{equation}
This is just a convolution of two Gaussian functions, $f$ and $g$, which is another Gaussian with mean equal to the sum of the means of $f$ and $g$ and variance equal to the sum of the variances of $f$ and $g$. The event likelihood is therefore
\begin{equation}
P(D_j | \sigma_{y_j}, \theta, M) \,=\, P(s_j | \theta_3, M) \ \frac{1}{\sqrt{2\pi (\sigma_{y_j}^2 + \omega^2)}} \, e^{-(y_j - \eta[s_j ; \theta_1])^2/2(\sigma_{y_j}^2 + \omega^2)} 
\label{eqn:eventlike4}
\end{equation}
i.e.\ involves no integration.  The time part of the time series model, $P(t_j=s_j | \theta_3, M)$, is simply evaluated at the measured time, $s_j$.  One particular application of this is to calculate the event likelihood for the {\em no-model}, the time series model in which there is no stochastic component and the deterministic component is just the mean of the signal. This is obtained by setting $\omega=0$ and $\eta = \overline{y_j}$ in equation~\ref{eqn:eventlike4}.   The signal is therefore expected to be just Gaussian noise fluctuations about a constant. The total likelihood for the no-model, $L^{\rm NM}$, is the product of these event likelihoods (equation~\ref{eqn:likelihood}).  This is a useful baseline model against which to compare the likelihood of other models. As this model has no adjustable parameters, this likelihood is equal to both the evidence and the K-fold CV likelihood.

\subsection{Both TSMod2 bypass and negligible time uncertainties}\label{sect:bypassanderrsmall}

If, in addition to a purely deterministic time series model, we also have 
negligible uncertainties on time, then the time part of the measurement model in equation~\ref{eqn:eventlike2} is a delta function, $\delta(t_j - s_j)$. The likelihood then involves no integration. 
If the signal part of the measurement model is a Gaussian, the likelihood is
\begin{equation}
P(D_j | \sigma_{y_j}, \theta, M) \,=\, P(s_j | \theta_3, M) \ \frac{1}{\sqrt{2\pi}\sigma_{y_j}} \, e^{-(y_j - \eta[s_j ; \theta_1] )^2/2\sigma_{y_j}^2} 
\label{eqn:eventlike5} \ \ .
\end{equation}
We also reach this result if we set $\omega=0$ in equation~\ref{eqn:eventlike4}.

\end{document}